\documentclass{jkas}

\def\year{2019} 
\def\month{March} 
\def\volume{00} 
\def\issue{0} 
\def\beginpage{1} 
\def\endpage{99} 
\def\received{---} 
\def\accepted{---} 
\date{Received \received; accepted \accepted}

\usepackage{blindtext}
\usepackage{caption}
\usepackage{rotating}
\usepackage[dvips]{color}

\title{
Core-Jet Blending Effects in Active Galactic Nuclei Under the Korean VLBI Network View at 43~GHz
}

\author[1,2]{Juan-Carlos~Algaba}
\author[3]{Jeffrey~Hodgson}
\author[3,4]{Sin-Cheol~Kang}
\author[2]{Dae-Won~Kim}
\author[5]{Jae-Young~Kim}
\author[3]{Jee~Won~Lee}
\author[3,4]{Sang-Sung~Lee}
\author[2]{Sascha~Trippe}

\affil[1]{Department of Physics, Faculty of Science, University of Malaya, 50603 Kuala Lumpur, Malaysia; \email{algaba@um.edu.my}}
\affil[2]{Department of Physics and Astronomy, Seoul National University, Gwanak-gu, Seoul 151-742, Korea}
\affil[3]{Korea Astronomy and Space Science Institute, 776, Daedeokdae-ro, Yuseong-gu, Daejeon, 305-348 Korea}
\affil[4]{Korea University of Science and Technology, 176 Gajeong-dong, Yuseong-gu, Daejeon, 305, 350, Korea}
\affil[5]{Max-Planck-Institut f\"ur Radioastronomie,  Auf dem H\"ugel 69, D-53121 Bonn, Germany}

\def\corrauthor{
J. C. Algaba
}

\def\runningauthor{
J. C. Algaba et al
}

\def\runningtitle{
Core-Jet Blending Effects in AGN Under the KVN View
}

\def\keywords{
galaxies: active -- methods: data analysis -- techniques: high angular resolution -- techniques: interferometric
}

\def\abstracttext{
A long standing problem in the study of Active Galactic Nuclei (AGNs) is that the observed VLBI core is in fact a blending of the actual AGN core (classically defined by the $\tau=1$ surface) and the upstream regions of the jet or optically thin emitting region flows. This blending may cause some biases towards the observables of the core, such as its flux density, size or brightness temperature, which may lead to misleading interpretation of the derived quantities and physics. We study the effects of such blending under the view of the Korean VLBI Network (KVN) for a sample of AGNs at 43~GHz by comparing their observed properties with observations with the Very Large Baseline Array (VLBA). Our results suggest that the observed core sizes are a factor $\sim11$ larger than these of VLBA, which is similar to the factor expected by considering the different resolutions of the two facilities. We suggest the use of this factor to consider blending effects in KVN measurements. Other parameters, such as flux density or brightness temperature, seem to possess a more complicated dependence.
}

\begin{document}
\jkashead 


\section{Introduction\label{sec:intro}}

The observable morphology of a typical radio-loud active galactic nuclei (AGN) consists of i) the core, an optically thick region classically defined by the $\tau=1$ surface, and ii) an optically thin jet or emitting region. In general, due to resolution limitations, the observed radio core is actually a blending of the actual AGN core and the upstream regions of the optically thin flows. This becomes very evident when studying the phenomenology of AGNs with instruments achieving further resolution limits: what was first conceived as the core region, can be now seen to contain further structure, consisting of a smaller core and additional emission or jet components. 

One of the long-standing problem in the study of these objects is that, even at resolutions of few milliarcseconds, provided by interferometric techniques such as VLBI, this core--jet blending effect can still be significant, as proven by observations by GMVA, VSOP or RadioAstron, which can resolve even further structure from what was considered to be the VLBI core \citep[see, e.g.,][]{Boccardi16,Asada16,Gomez16}. This blending effect can greatly affect the observables, such as polarization, core size, or core shift frequency dependence, as contribution from both the core and the innermost jet regions are integrated together. Thus, such blending has to be treated carefully and an analysis has to be done to properly understand its effects on our observable quantities.

Various methods have been used in the literature to consider such blending effects and its implications in the data analysis. One approach is to consider observations with better resolution and to compare the observables, such as the morphology or flux density, in order to understand how the different resolution may affect these \citep[e.g.][]{Kovalev08,Pushkarev12}. Although better resolutions can be obtained with increasing frequency, physical properties of the source may also be different at various frequency (due to e.g., opacity effects), and thus different arrays or, if possible, different array configurations observing at the same frequency are preferred for such analysis. A different approach would consist of the convolution of a pre-existing high resolution map with a larger beam size or the flagging of data at various UV-distances in the interferometric UV-plane to simulate a map of lower resolution \citep[e.g.][]{Hovatta14}. A third approach would include the analysis of Monte Carlo simulations on a predefined model \citep[e.g.][]{Mahmud13}.  

The Korean VLBI Network (KVN) is a unique interferometric array located in the Korean peninsula. Consisting of three 21-m antennas equipped with a multi-receiver band system, it can observe at 22, 43, 86 and 129~GHz, higher frequencies than most other VLBI networks, simultaneously. With baselines between 305--476~km, achievable resolutions at these frequencies can reach about $\gtrsim1$~mas, which makes the KVN an excellent tool to resolve the innermost regions of AGNs. Nonetheless, as we have mentioned, for a robust analysis, it is necessary to first consider the blending effects that can be expected from the KVN view.

A pioneering work where KVN source blending issues is discussed is that of \cite{Rioja14}. Although their work mainly focuses on astrometric issues, they include an analysis of the source structure effects in the KVN, including structure blending effects as compared with both high- and matched-resolution VLBA images. As they find, this blending has a large impact in astrometric measurements, becoming the dominant source of errors in astrometric measurements of extended sources. Its magnitude, however, seems to be different case by case, suggesting that a large sample should be needed for  proper study. 

In this paper we study the core properties, such as the core size or core brightness temperature, of several AGNs observed with the KVN and compare them with observations with the Very Large Baseline Array (VLBA), an array which offers better angular resolution. In this way, we estimate the core-jet blending and its effects on such observables. \cite{Kim19} will do parallel analysis using a different approach. The paper is organized as follows: in Section 2 we describe the observations used for our analysis. In Section 3 we show our results and comparison. In Section 4 we discuss these and possible implications. A summary can be found in Section 5.

\section{Observations}

\subsection{Data Selection Criteria}

In order to study possible core blending effects, we intend to compare our KVN measurements with other VLBI observations of the same source capable of clearly resolving components or features upstream of the jet that may be blended with the KVN core. Given the strong variability of these objects, with timescales of even just a few days \citep[see e.g.][]{Wagner95}, comparison should ideally be performed with (\mbox{quasi-})simultaneous observations. For consistency and robustness of the results, one should ideally include various epochs, not too sparse in time, if possible. 

However, observations performed (quasi-) simultaneously in different arrays are not always possible, except for very particular cases where the typical VLBI dynamic scheduling is superseded by a strong science case very particular source or event scenarios. An alternative approach to circumvent these limitations is to investigate data obtained for an extended period of time and, if the relative observed variability is not significant, consider the mean values of the observables. 

To analyze the KVN data, we used data from the iMOGABA \citep[interferometric monitoring of gamma ray bright AGNs; see][]{Lee13,Algaba15,Lee16} program, which observes a total of 30 well known AGNs with a mean cadence of about a month at 22, 43, 86 and 129~GHz. This is, to date, the best source for AGN monitoring with the KVN array. Unfortunately, there are very little VLBI multi--epoch observations or monitoring programs available for comparison. In particular, the authors are not aware of any VLBI program at 22 or 86~GHz with good cadence. Similarly, the situation at 129~GHz is very tricky since this is not a common VLBI observing frequency. Only at 43~GHz the Boston University VLBA-BU-BLAZAR Program provides an excellent systematic monitoring of AGNs.

Consequently, in this paper we will focus on the comparison of KVN iMOGABA 43~GHz data with the VLBA-BU-BLAZAR program which, thanks to the much larger baselines, provides a resolution better by a factor of $\sim18$. Although not complete in terms of frequency space nor in the framework of source characteristics, this will provide a fundamental first step test to understand the KVN core blending effects. With this in mind, we will leave the analysis of other KVN frequencies for a forthcoming study.

One of the most immediate observable that gets affected by blending effects is the core size. Indeed, as the observed VLBI core is a combination of the actual $\tau=1$ surface (the actual core) and the innermost unresolved regions of the jet, blending plays a significant role in its resulting observed size. The larger the innermost jet regions merged with the core due to blending, the larger the apparent observed size of the core will be. 
Being one of the more direct quantities that can be easily measured in VLBI observations, the core size seems like an ideal proxy to consider the blending effects. Similarly, the flux density is a very straightforward observable that can also be affected by the area measured. Finally, the brightness temperature combines these two factors and has proven to be a quantity of great importance in high resolution mapping \citep[see e.g.,][]{Bruni17,Kardashev17,Pilipenko18}. Thus, in this work, we will consider these three different observables.

\subsection{The Data}

Information of the data obtained from the VLBA-BU-BLAZAR program, including model-fitting and properties of the VLBA core flux density and size is summarized in \cite{Jorstad17}. We note however that, although this program is still ongoing,  \cite{Jorstad17} limits its analysis to epochs prior June 2013. In some cases, we were able to access the later public data and continue the model--fit of the source for subsequent epochs \citep[such as for e.g., 1633+382; see][for further details]{Algaba18a,Algaba18b}. The VLBA-BU-BLAZAR program does not contain information about M87. For this source, we used VLBA data from \cite{Hada13}, which contains a total of 7 epochs observations at 43~GHz, once upper limits of the core size are excluded from the analysis. No VLBI core flux densities are shown in this paper, but we used a fiducial value of 0.7~Jy \citep{Ly07,Walker18}.

Regarding the KVN data from the iMOGABA program, only a handful of iMOGABA sources have currently been already analyzed in depth. Other sources are still being investigated or under analysis. Data for 0716+714 is publicly available in \cite{LeeJW17}; data for 1156+295 is described in \cite{Kang18}; data for 1633+382 is summarized in \cite{Algaba18a,Algaba18b}; data from M87 can be inspected in \cite{KimJY18}; and data for BL~Lac is investigated in \cite{KimDW17}.  In order to obtain state-of-the-art information regarding the properties of the rest of the sources, an iMOGABA modelf-fitting \textsc{Difmap} script has been implemented \citep{Hodgson16}. In a nutshell, this script finds the best model based on one circular gaussian to fit the core. This is expected to work well given that  iMOGABA sources at 43~GHz are mostly either point-like or core-dominated. KVN uncertainties for 43 GHz should be close to 10\%. A detailed discussion is provided in \cite{Lee16}.

\begin{figure*}
\includegraphics[scale=0.55,trim={0cm 0cm 0cm 0cm},clip]{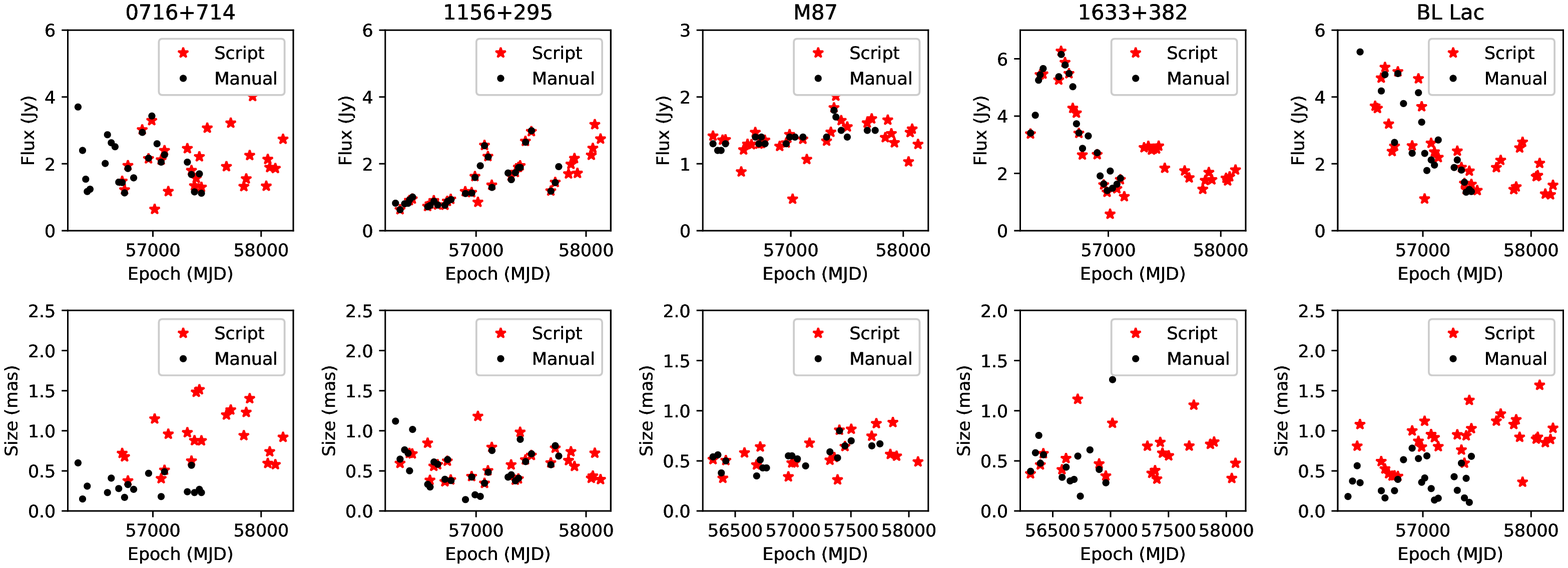}
\caption{Comparison of the flux densities (top) and core sizes (bottom) obtained using the iMOGABA model fitting script (red stars) and manual fitting (black dots).} 
\label{scriptVSmanual}
\end{figure*}

In Figure \ref{scriptVSmanual} we show the results obtained with this script compared with the bona--fide more robust manual analysis which may include, in some cases, additional components. It is clear that, considering their uncertainties, the flux densities obtained with the script are quite reliable and follows well these obtained with a more careful analysis. The core sizes seem to also roughly match, except for the cases of extended structures with significant flux, such as that of  0716+714 or BL~Lac, where the script overestimates the size. Nonetheless, such difference is only by a factor of $\lesssim2$ at most, which is not dramatic for our study and, as mentioned earlier, will happen on only very few cases. We thus consider that, in a statistical sense, the script works well for our purposes here.

\section{Results}
Flux densities and core sizes were obtained for 25 sources. Brightness temperatures were calculated using the relationship $T_b=1.22\times10^{12}S(1+z)/43^2/d^2$, where $S$ and $d$ are the model fitted core flux densities and core sizes. Compiled data can be examined in Figure \ref{fig_alldata}. In general, we were not able to obtain (quasi-) simultaneous data, and there is a gap between VLBI and KVN data for most of the sources, except for the case of 1633+382 and 1156+295. It seems however that, whilst certain variability, inherent to these sources, is still clear, its dispersion in terms of the quantities observed here is not too critical for our purposes. In fact, it seems that the dispersion is of the same order of the respective measured quantities, if not smaller. On the other hand, comparison between the KVN and VLBI arrays show that, whereas the measured flux densities appears to be quite similar for many sources, measured core sizes and brightness temperatures are different in the two arrays by at least an order of magnitude. Table \ref{medianquantities} summarizes the median values of flux density and core size calculated among all sources and their dispersion in a quantitative manner.

\begin{table}
\caption{Sample Median Quantities\label{medianquantities}}
\centering
\begin{tabular}{cccc}
\toprule
Array   & Flux (Jy) & Size (mas) & $T_b$ (K)  \\
\midrule
VLBA & $1.2\pm0.7$ & $0.06\pm0.02$ & $(2.5\pm2.6)\times10^{11}$\\
KVN & $2.1\pm1.3$ & $0.77\pm0.27$ & $(4.1\pm4.5)\times10^{9}$\\
\bottomrule
\end{tabular}
\end{table}

\begin{figure}
\includegraphics[scale=0.58,trim={0cm 0cm 0cm 0cm},clip]{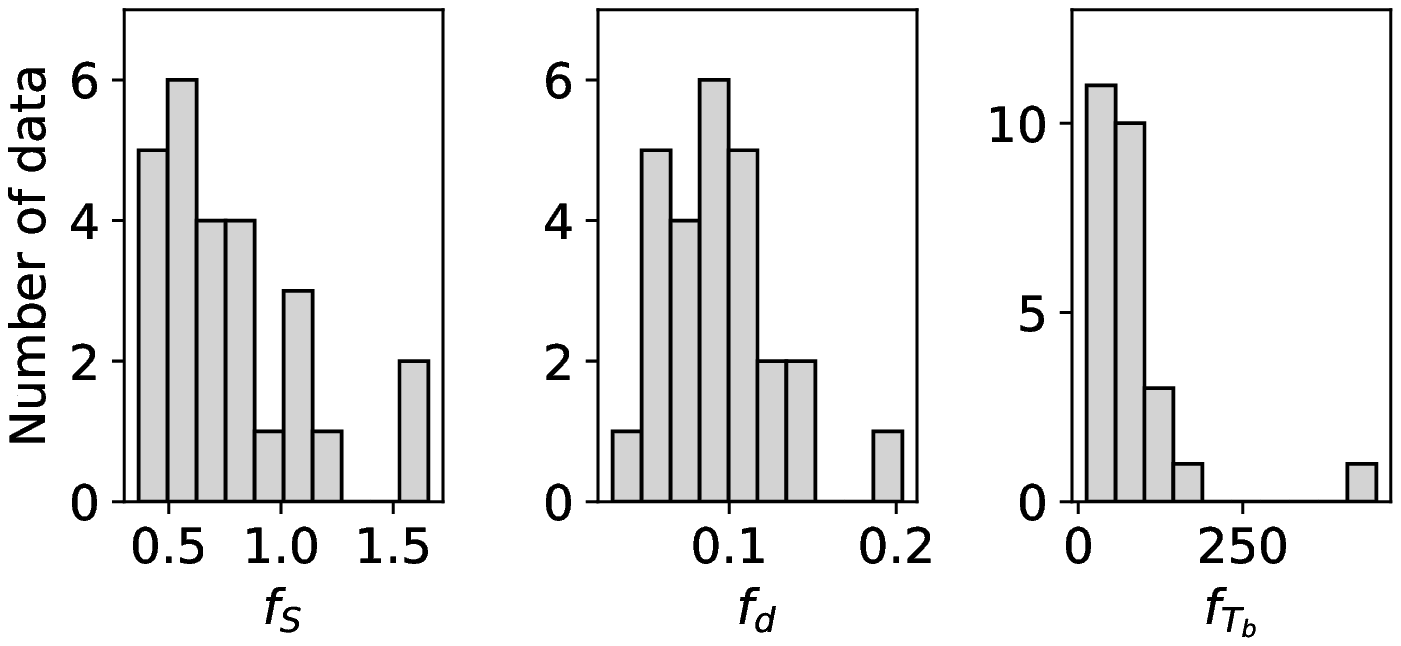}
\caption{Histograms for the measured fractional quantities $f_S, f_d$ and $f_{T_b}$.} 
\label{fig_histogram}
\end{figure}

In Table \ref{table} we summarize some relevant source properties and the average quantities obtained from our data. Columns 1, 2 and 3 show the source name, its redshift and it viewing angle, from \cite{Hovatta09}. Columns 4, 5 and 6 indicate the core flux $S$ under the VLBA and KVN perspectives, and a compactness factor, $f_S=S^{VLBA}/S^{KVN}$. Columns 7, 8 and 9 show the core size $d$ under the VLBA and KVN perspectives, and the core size ratio $f_d=d^{VLBA}/d^{KVN}$. Similarly, columns 10, 11 and 12 show the core brightness temperature $T_b$ under the VLBA and KVN perspectives, and the core brightness temperature ratio $f_{T_b}=T_b^{VLBA}/T_b^{KVN}$. Statistical values for the fractional quantities are shown in Table \ref{medianfractional}, and a histogram is shown in Figure \ref{fig_histogram}.

\begin{figure*}

\includegraphics[scale=0.4,trim={0.0cm 0cm 1.2cm 0cm},clip]{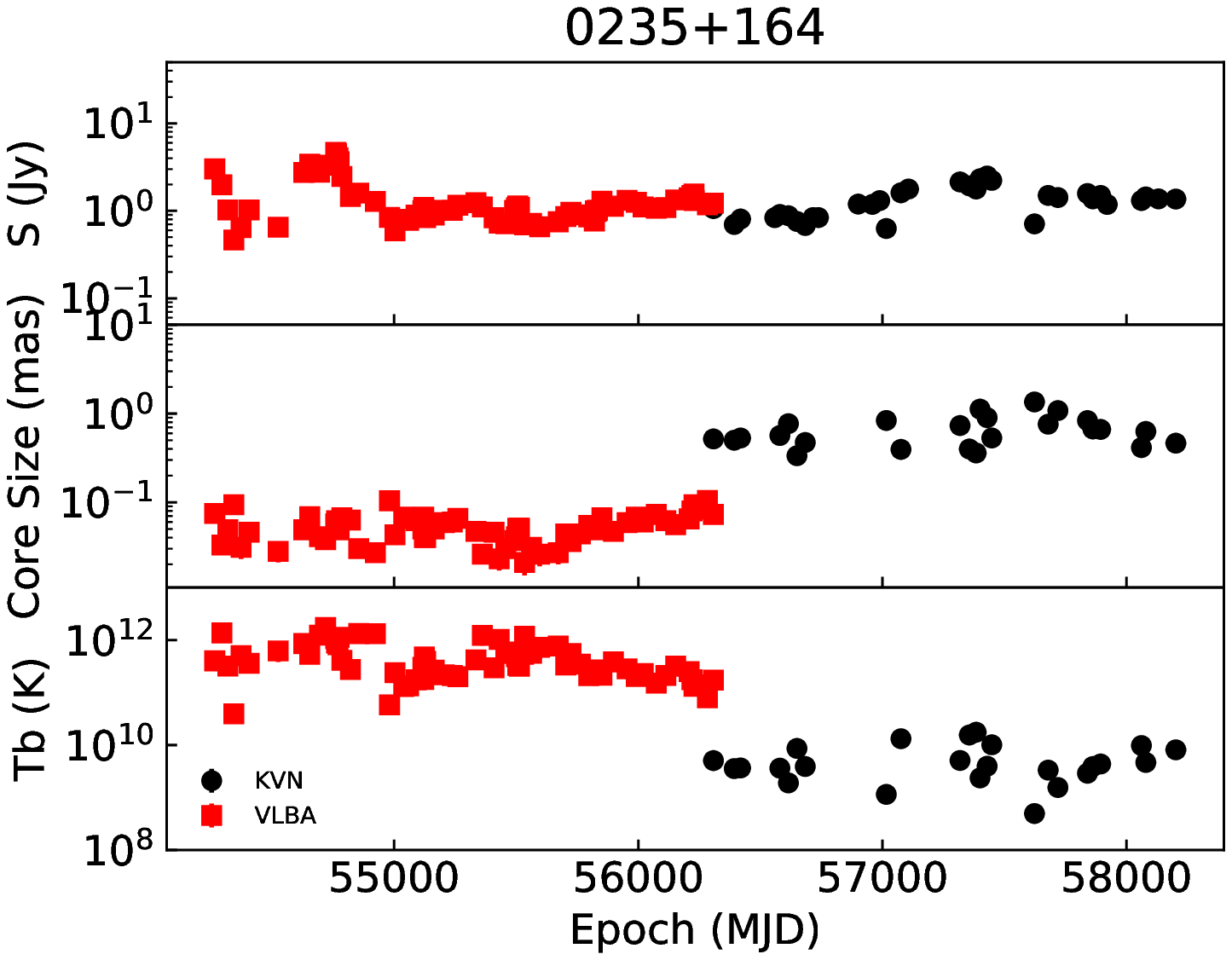}
\includegraphics[scale=0.4,trim={0.0cm 0cm 1.2cm 0cm},clip]{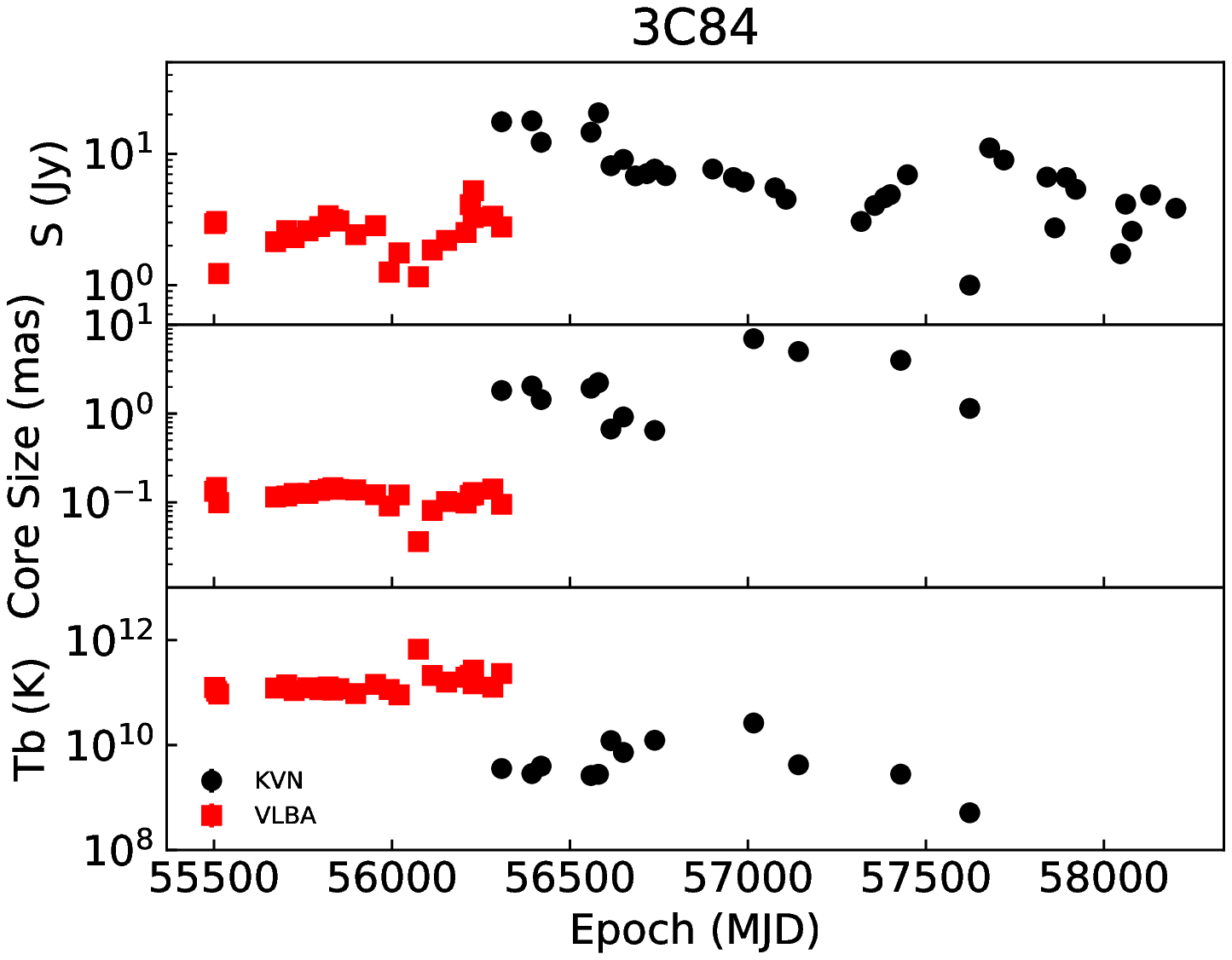}
\includegraphics[scale=0.4,trim={0.0cm 0cm 1.2cm 0cm},clip]{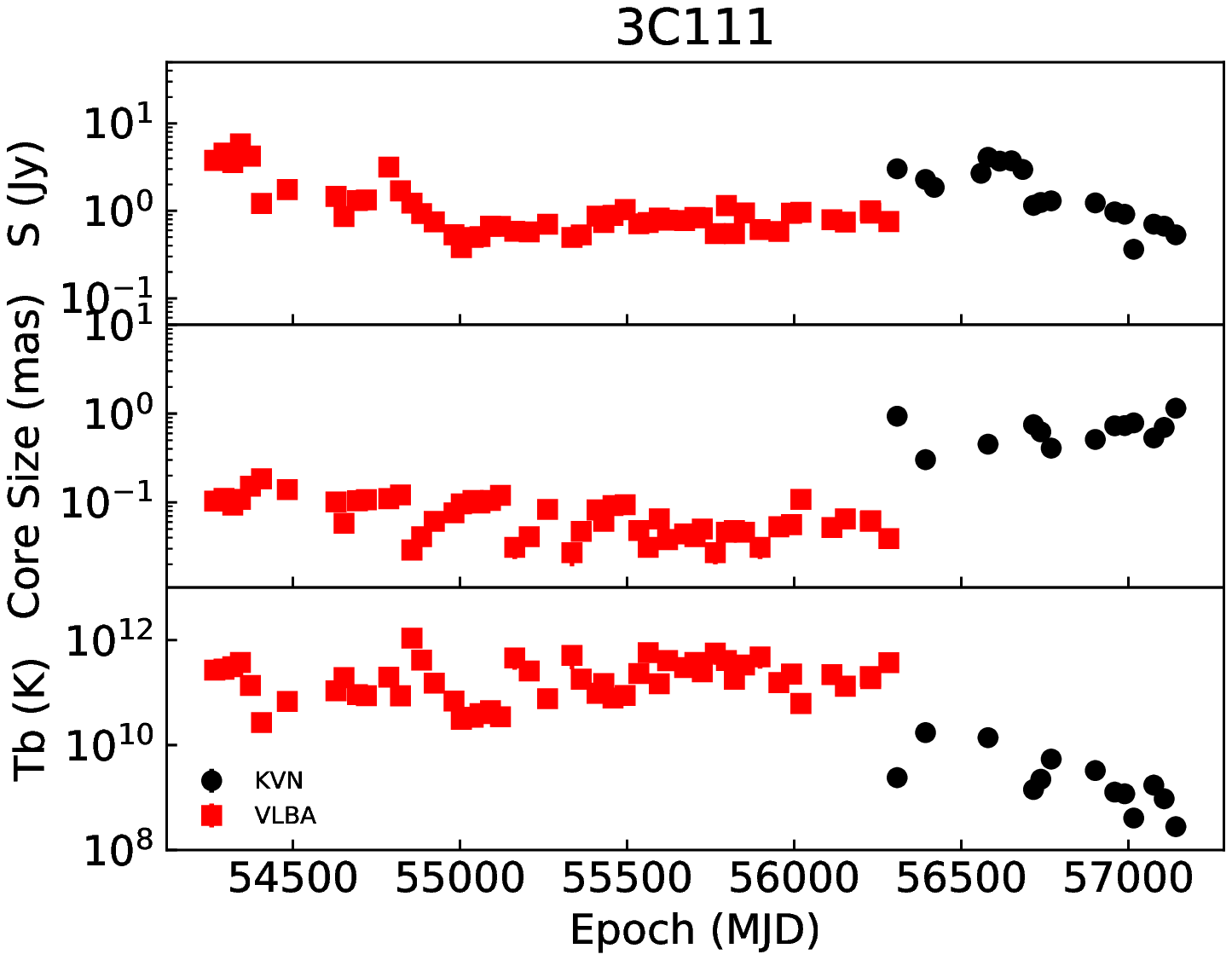}\\
\includegraphics[scale=0.4,trim={0.0cm 0cm 1.2cm 0cm},clip]{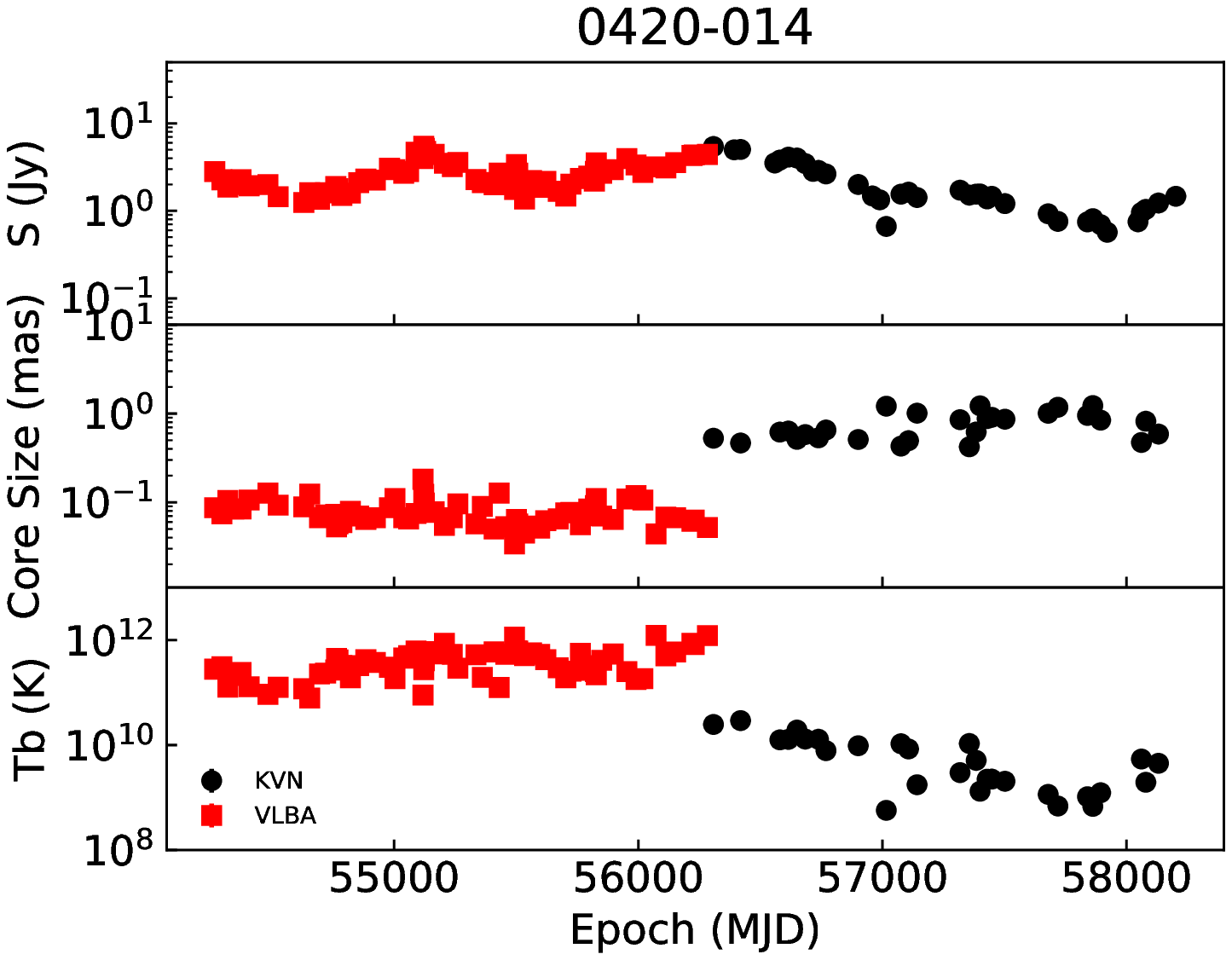}
\includegraphics[scale=0.4,trim={0.0cm 0cm 1.2cm 0cm},clip]{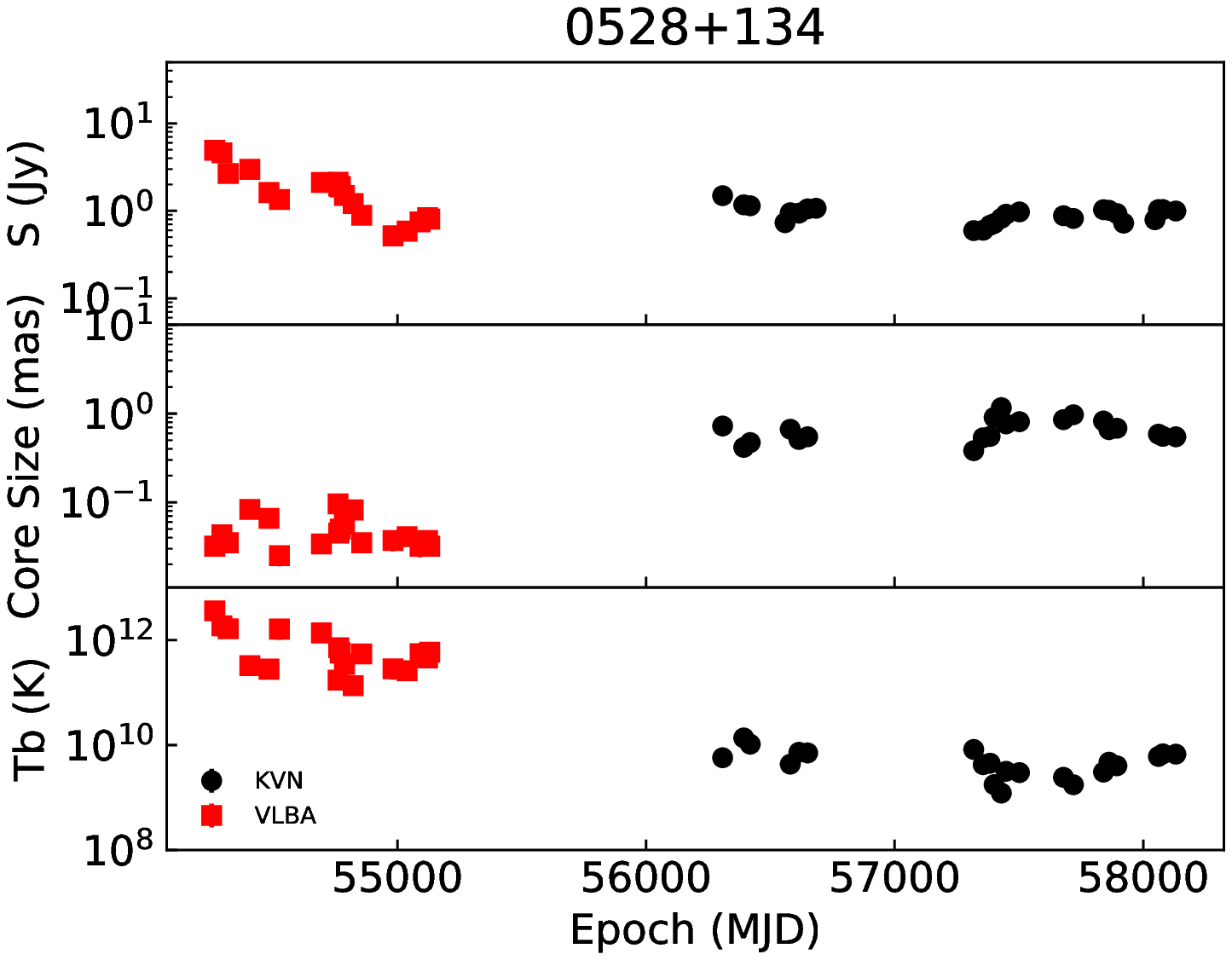}
\includegraphics[scale=0.4,trim={0.0cm 0cm 1.2cm 0cm},clip]{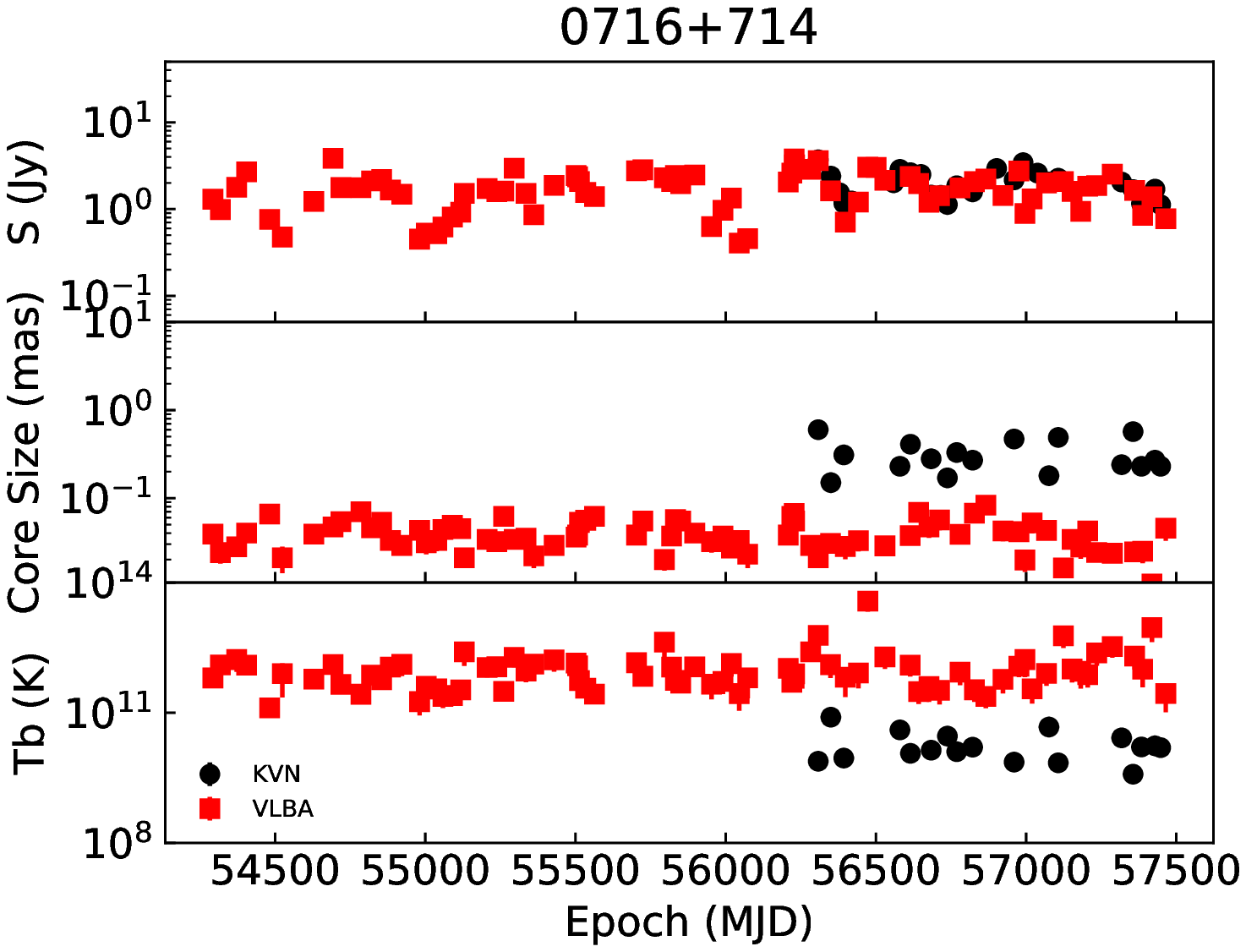}\\
\includegraphics[scale=0.4,trim={0.0cm 0cm 1.2cm 0cm},clip]{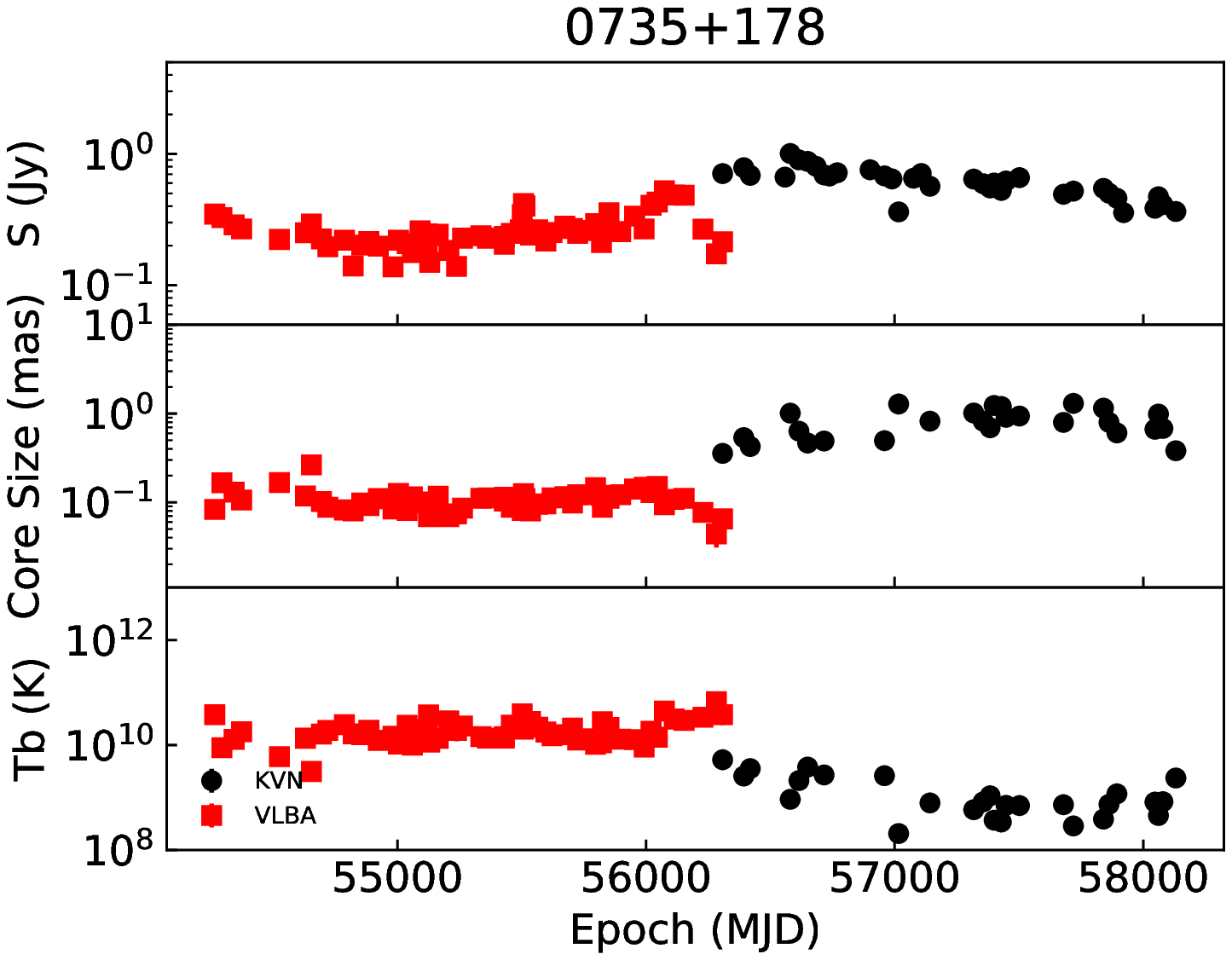}
\includegraphics[scale=0.4,trim={0.0cm 0cm 1.2cm 0cm},clip]{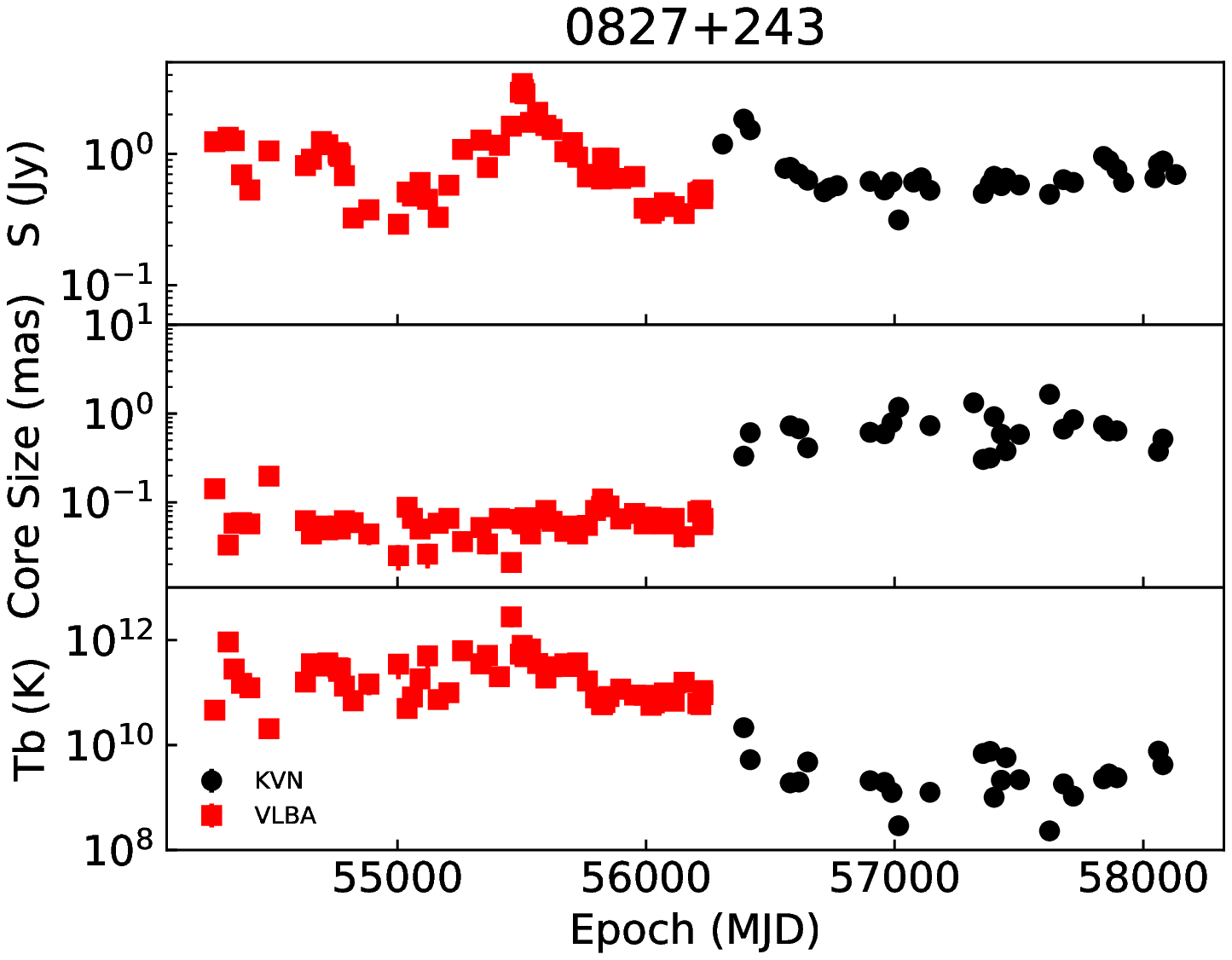}
\includegraphics[scale=0.4,trim={0.0cm 0cm 1.2cm 0cm},clip]{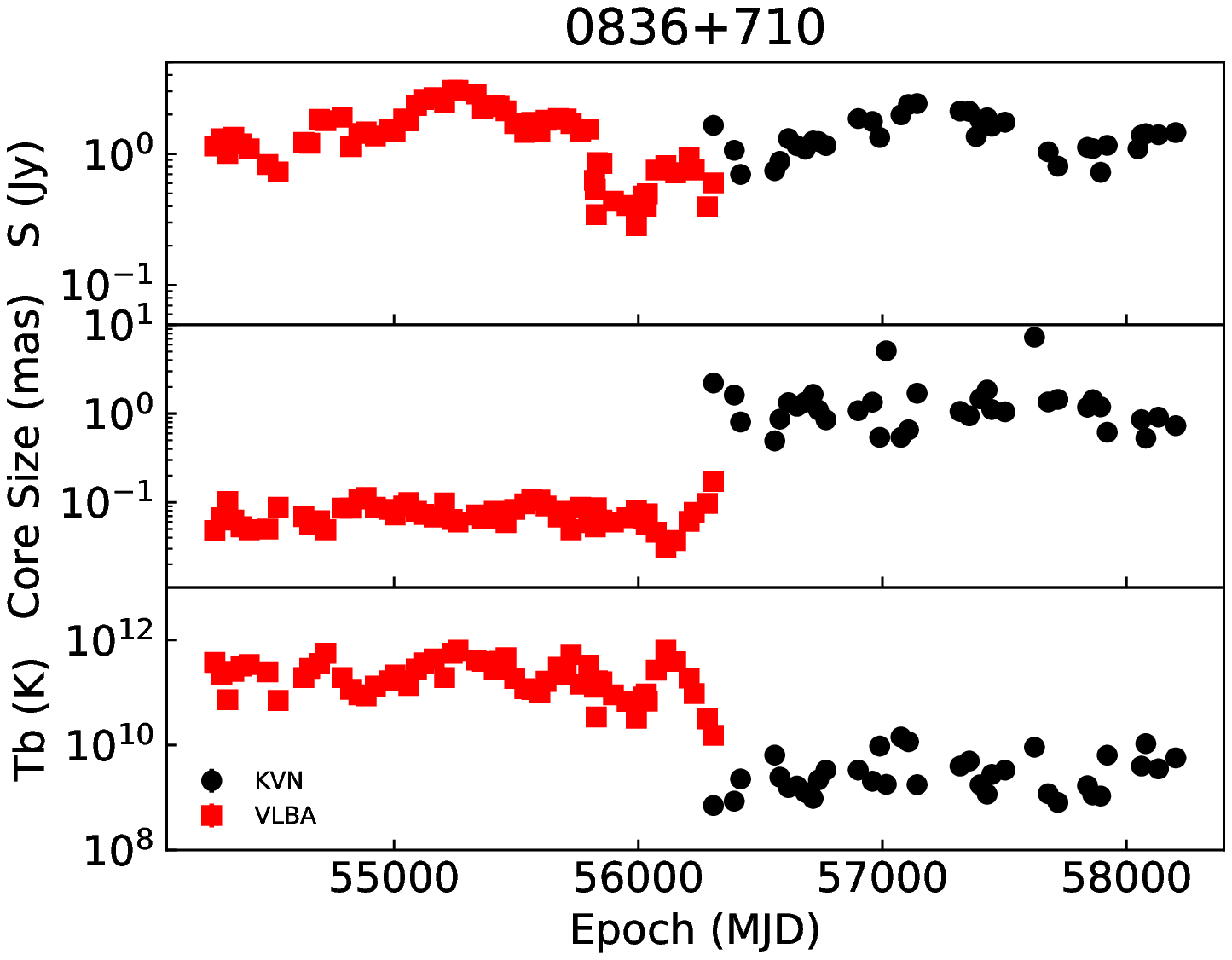}\\
\includegraphics[scale=0.4,trim={0.0cm 0cm 1.2cm 0cm},clip]{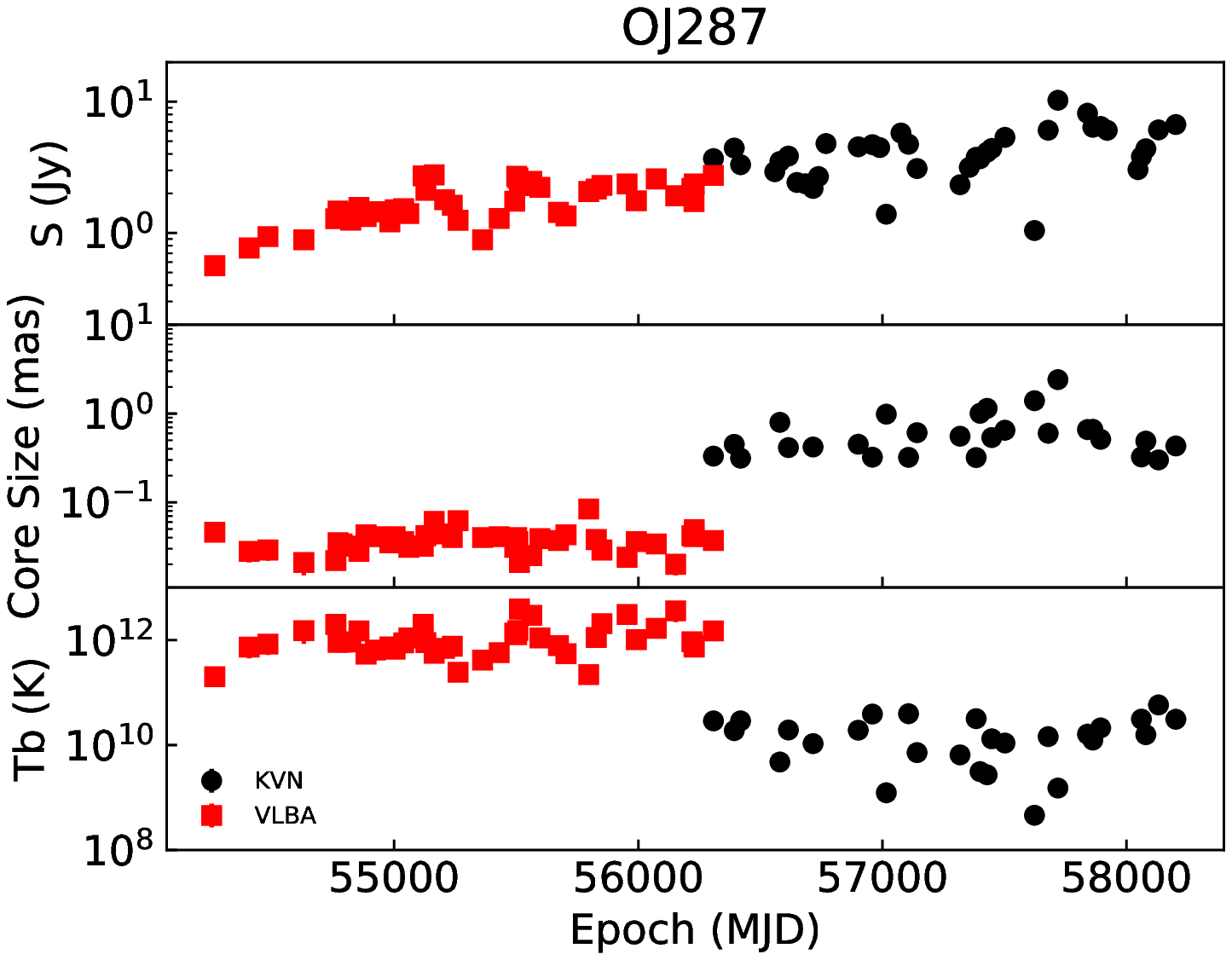}
\includegraphics[scale=0.4,trim={0.0cm 0cm 1.2cm 0cm},clip]{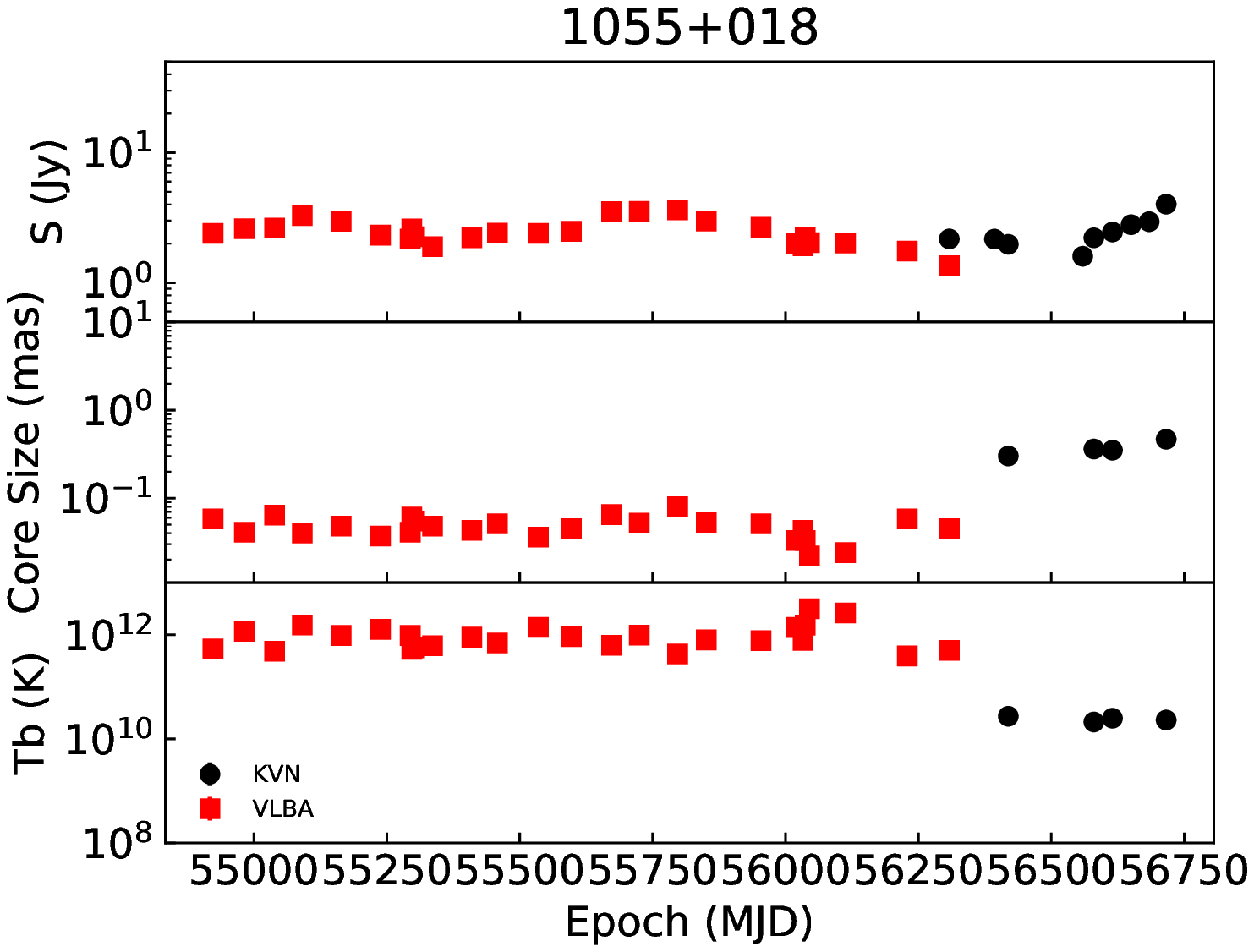}
\includegraphics[scale=0.4,trim={0.0cm 0cm 1.2cm 0cm},clip]{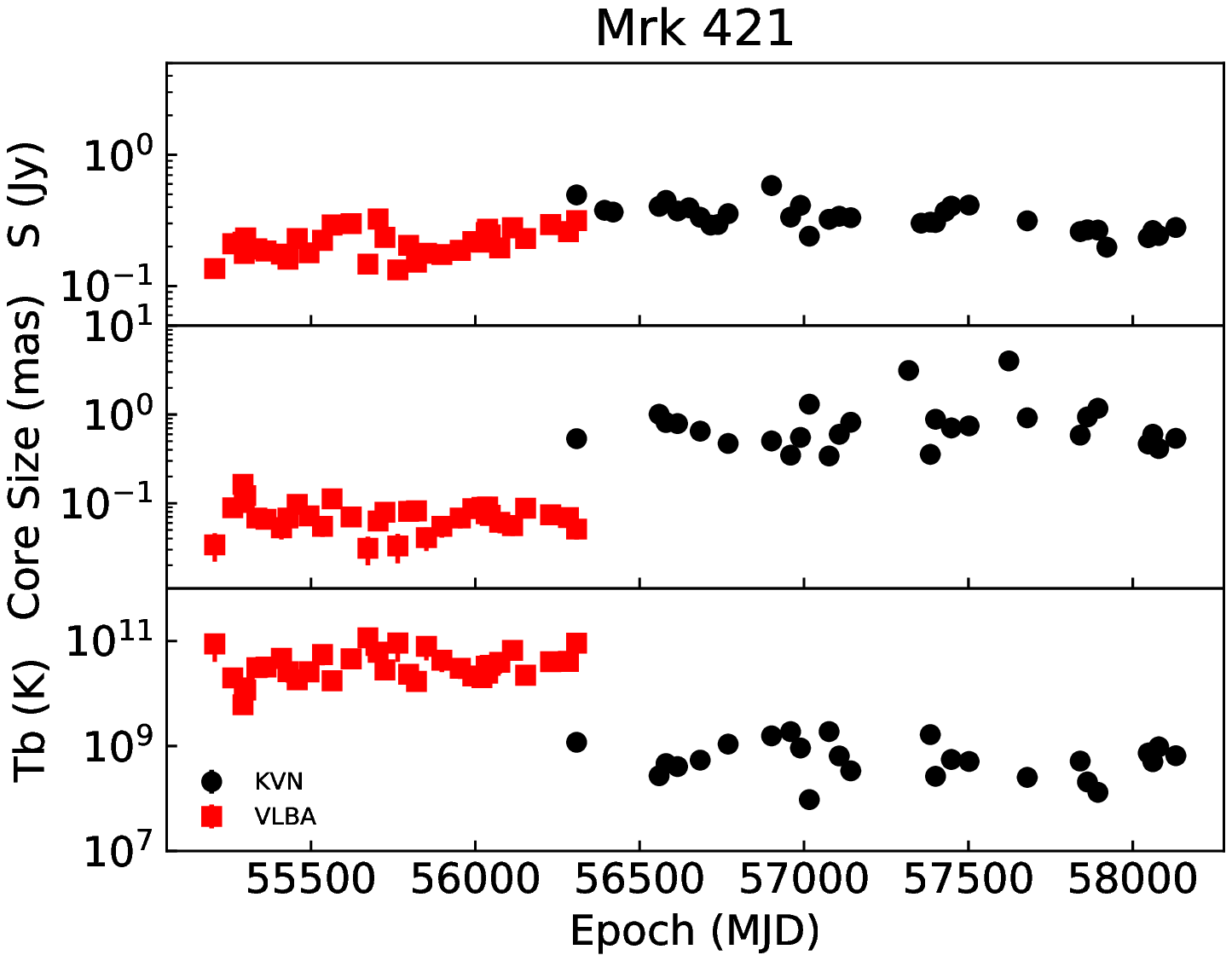}\\
\includegraphics[scale=0.4,trim={0.0cm 0cm 1.2cm 0cm},clip]{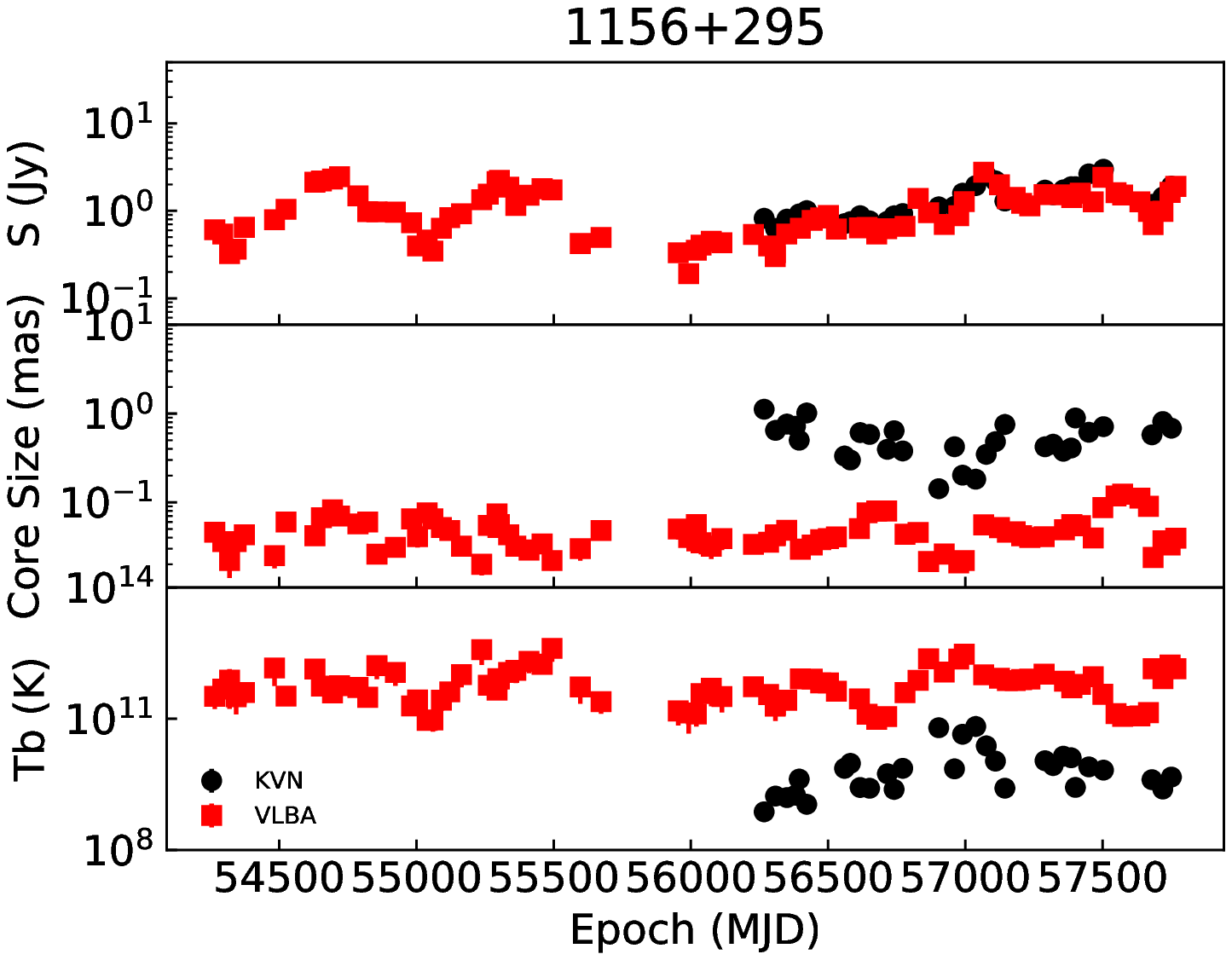}
\includegraphics[scale=0.4,trim={0.0cm 0cm 1.2cm 0cm},clip]{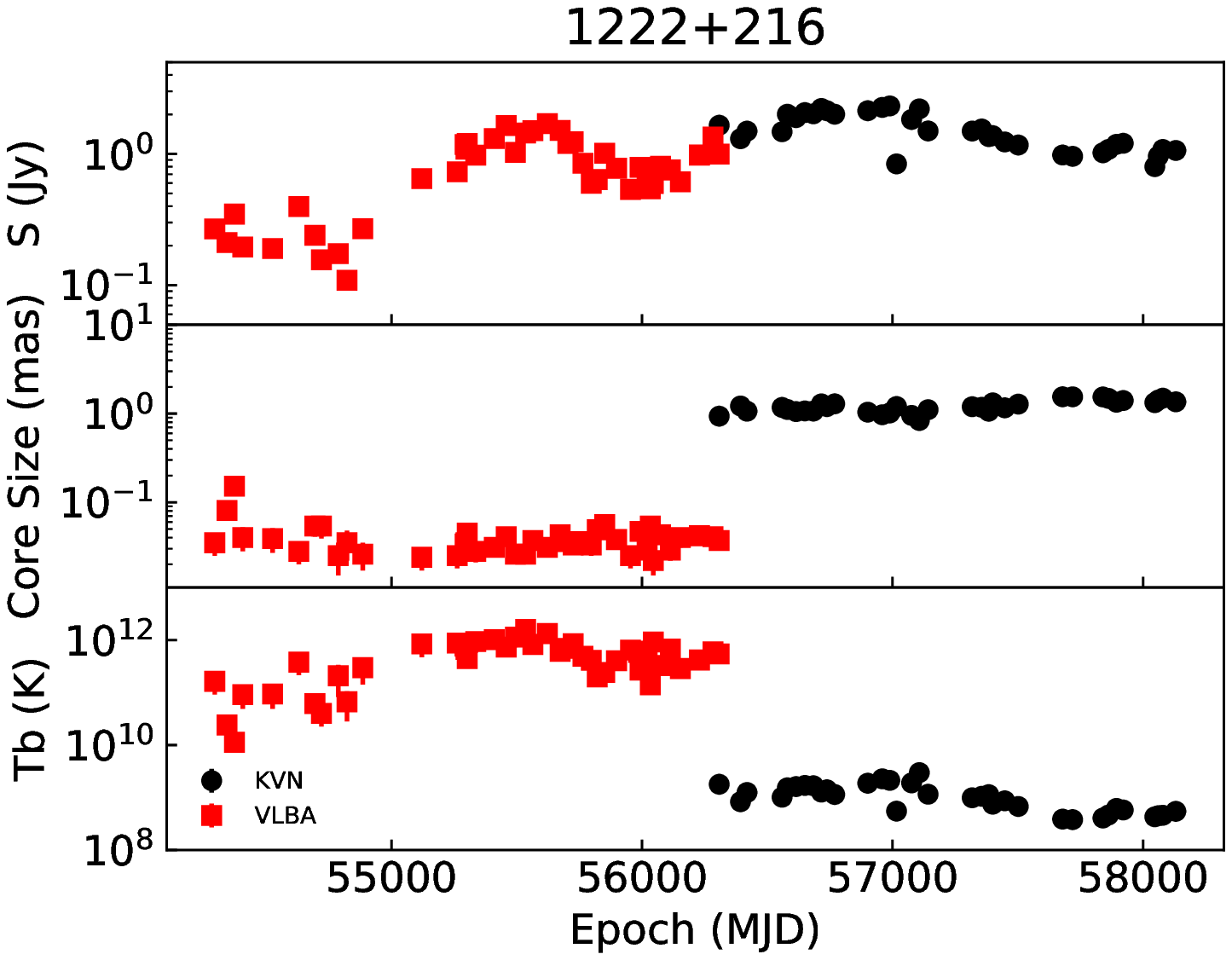}
\includegraphics[scale=0.4,trim={0.0cm 0cm 1.2cm 0cm},clip]{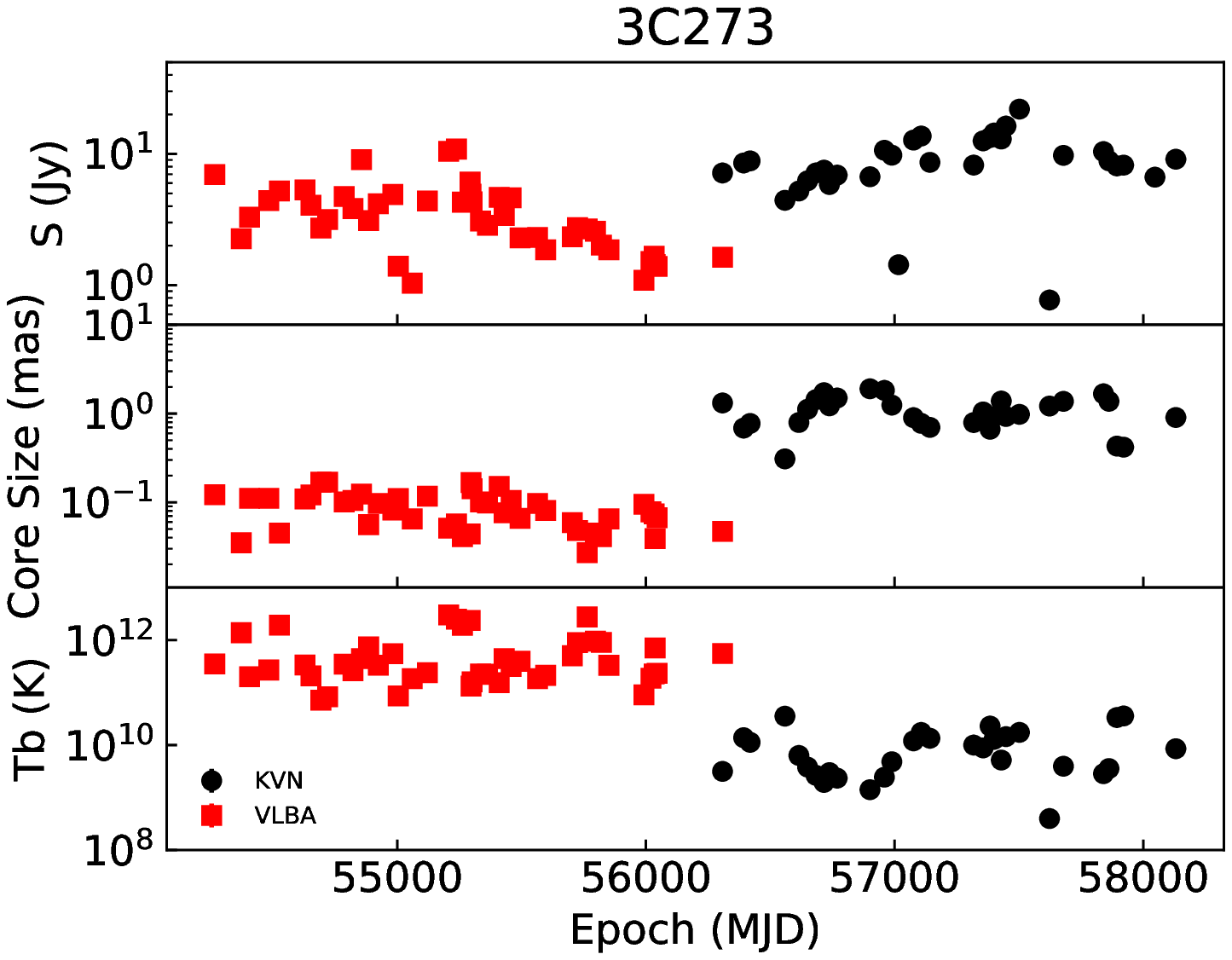}\\
\caption{Comparison of core sizes and brightness temperatures observed with VLBA (red squares) and KVN (black circles).} 
\label{fig_alldata}
\end{figure*}
\begin{figure*}\ContinuedFloat
\includegraphics[scale=0.4,trim={0.0cm 0cm 1.2cm 0cm},clip]{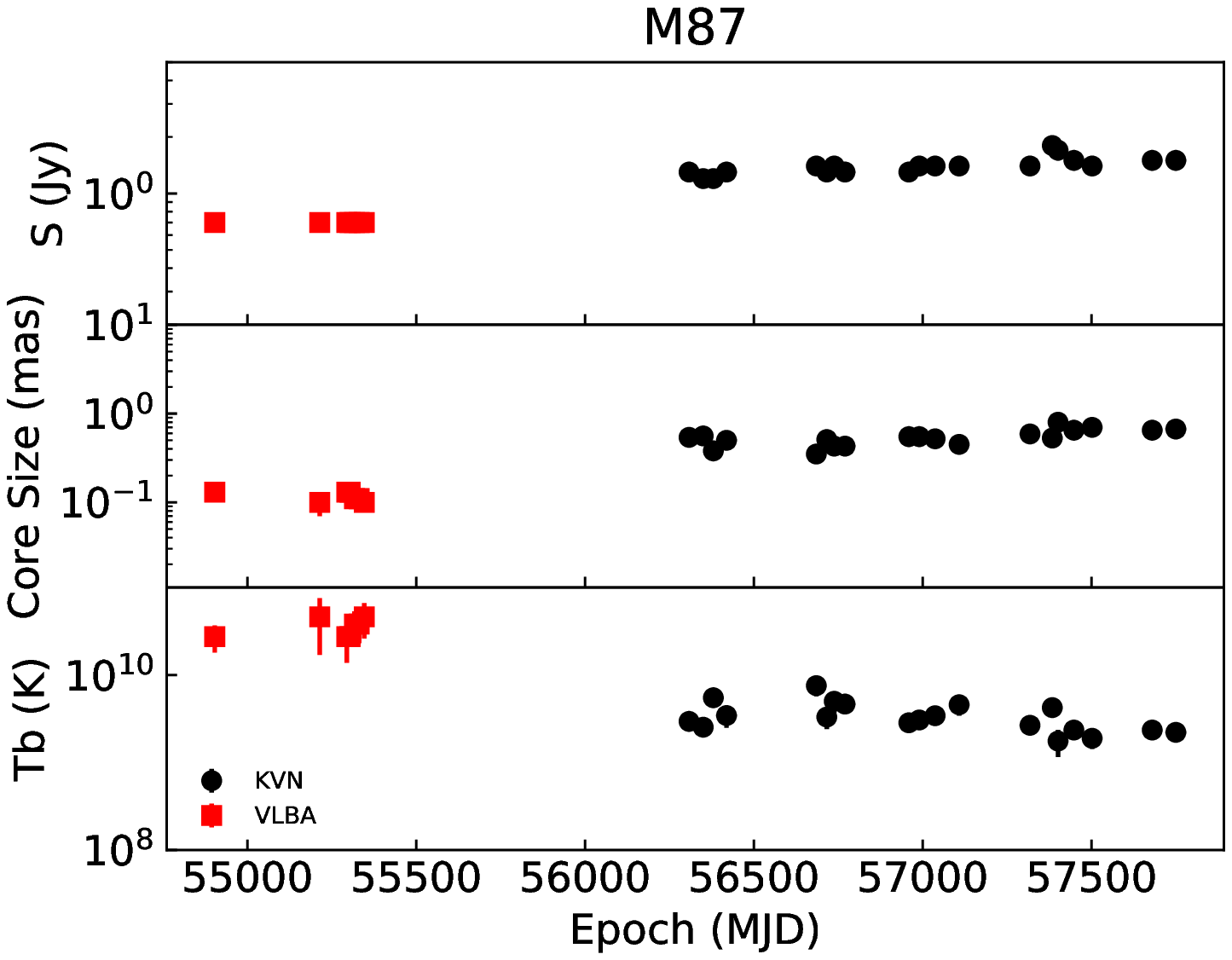}
\includegraphics[scale=0.4,trim={0.0cm 0cm 1.2cm 0cm},clip]{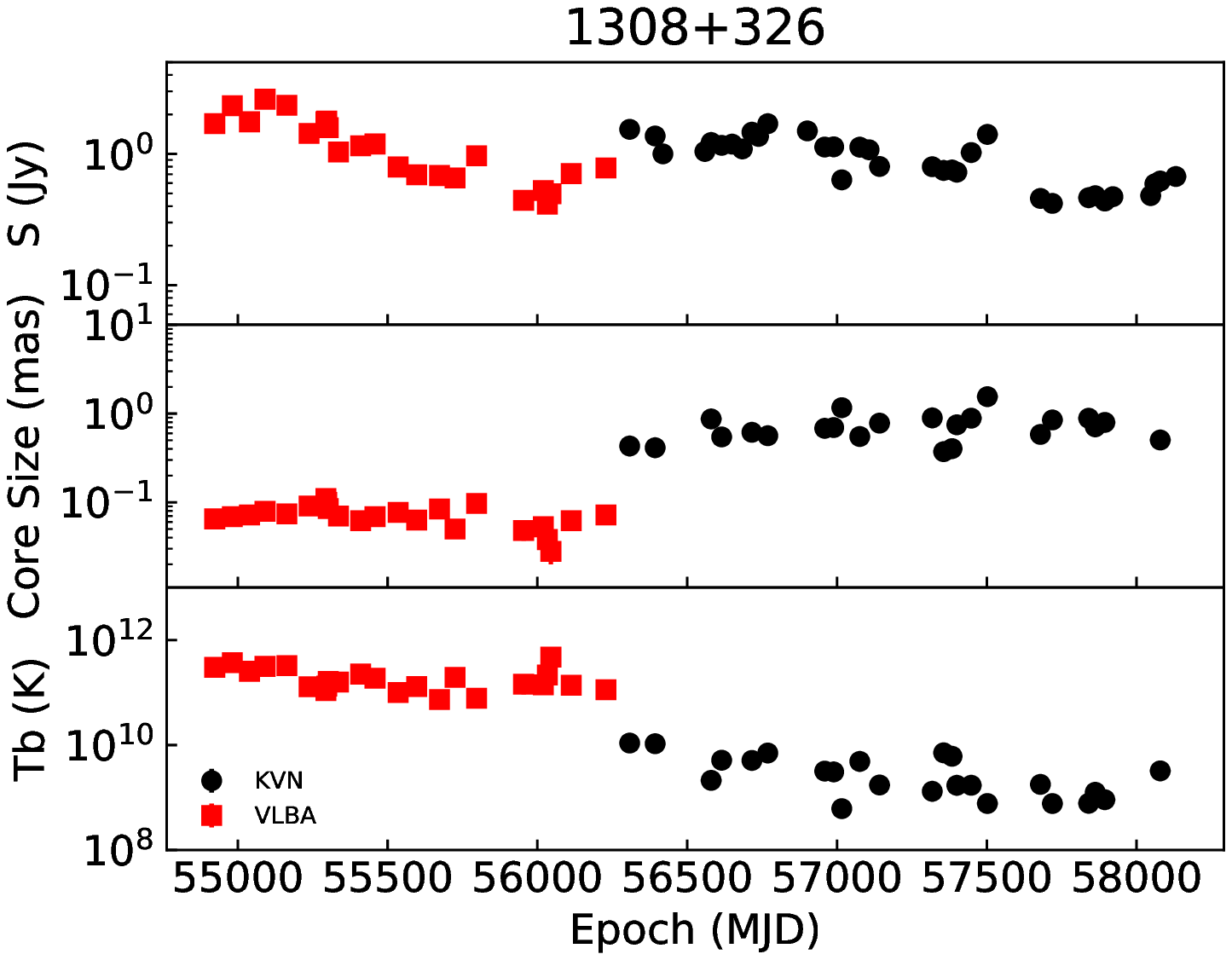}
\includegraphics[scale=0.4,trim={0.0cm 0cm 1.2cm 0cm},clip]{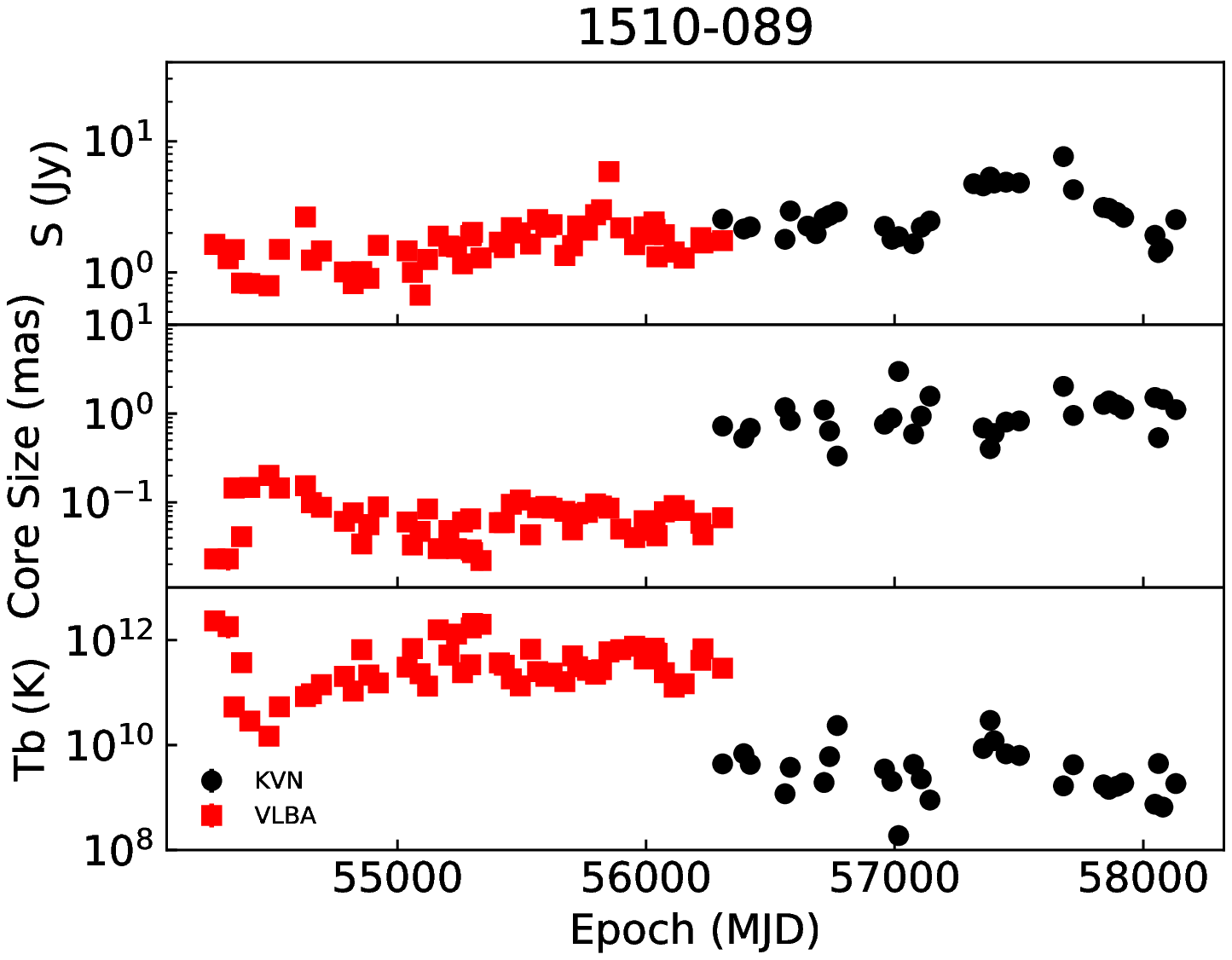}\\
\includegraphics[scale=0.4,trim={0.0cm 0cm 1.2cm 0cm},clip]{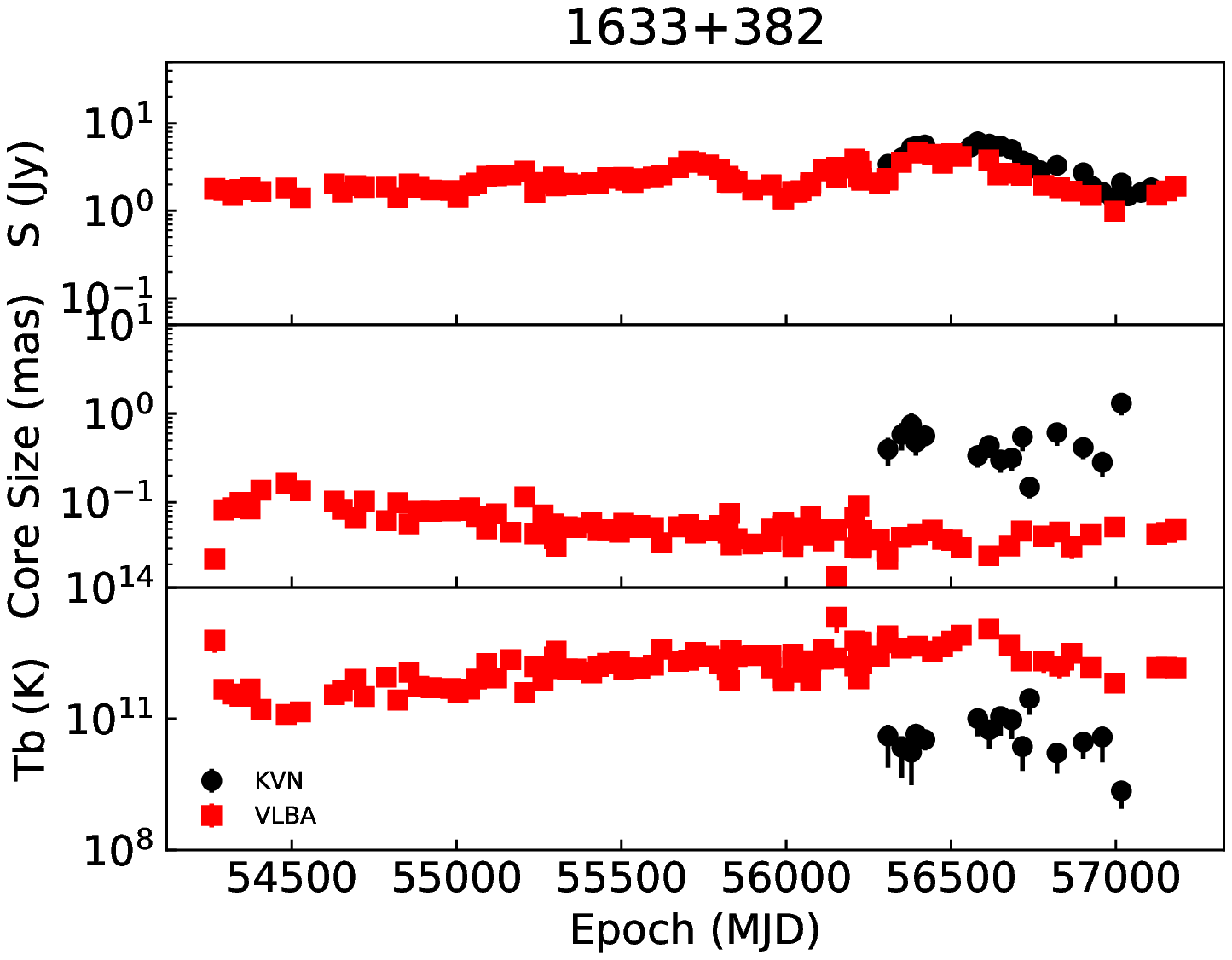}
\includegraphics[scale=0.4,trim={0.0cm 0cm 1.2cm 0cm},clip]{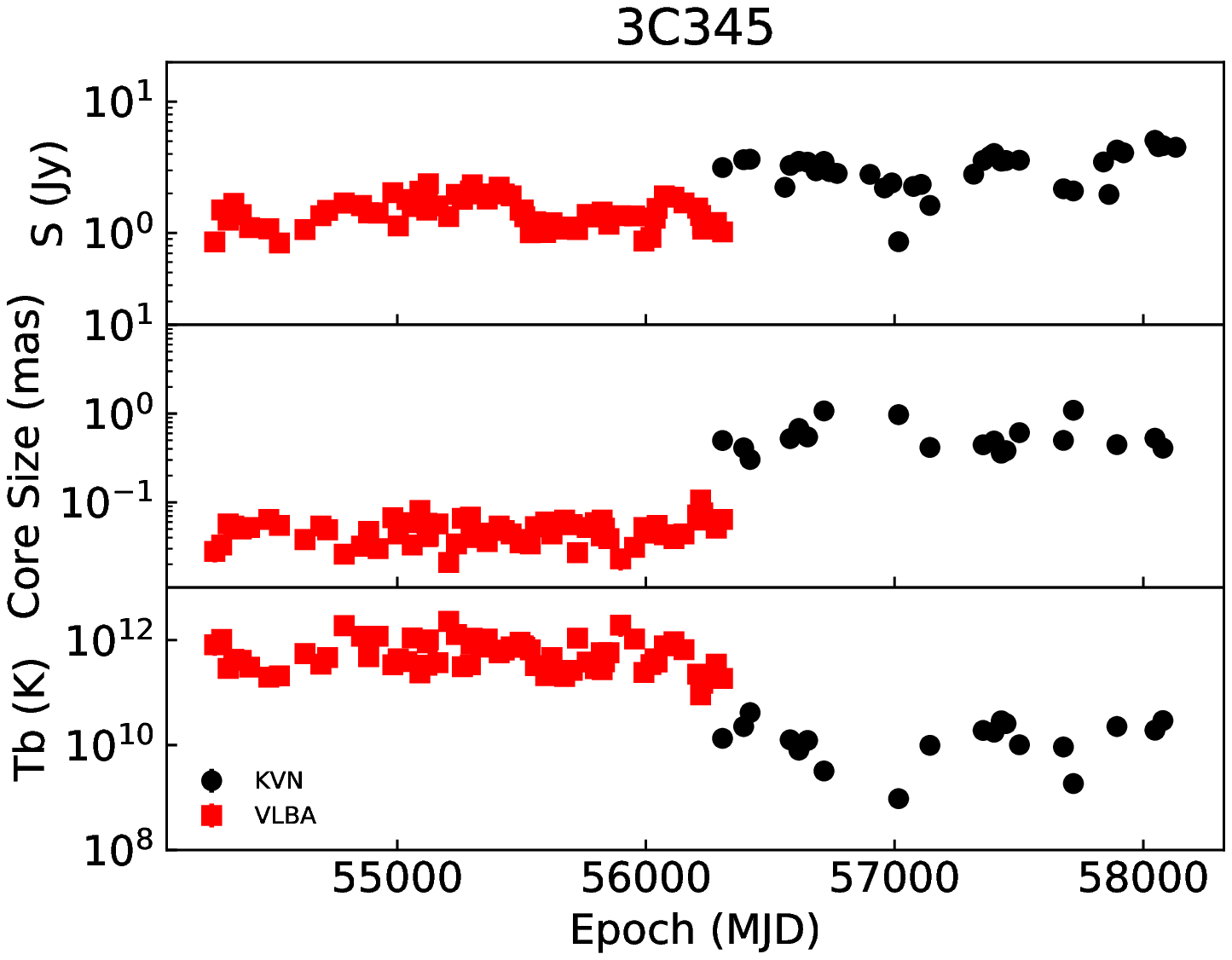}
\includegraphics[scale=0.4,trim={0.0cm 0cm 1.2cm 0cm},clip]{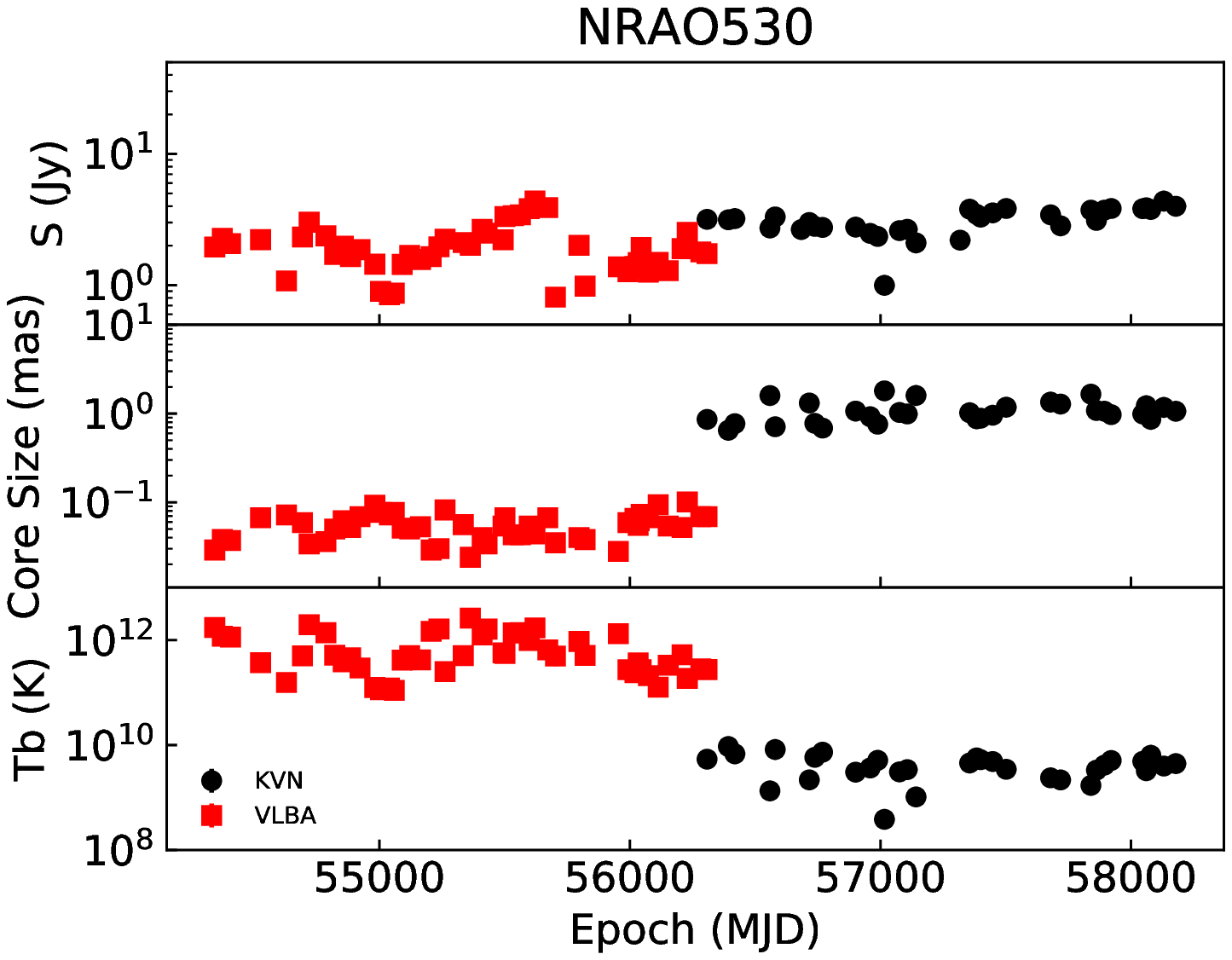}\\
\includegraphics[scale=0.4,trim={0.0cm 0cm 1.2cm 0cm},clip]{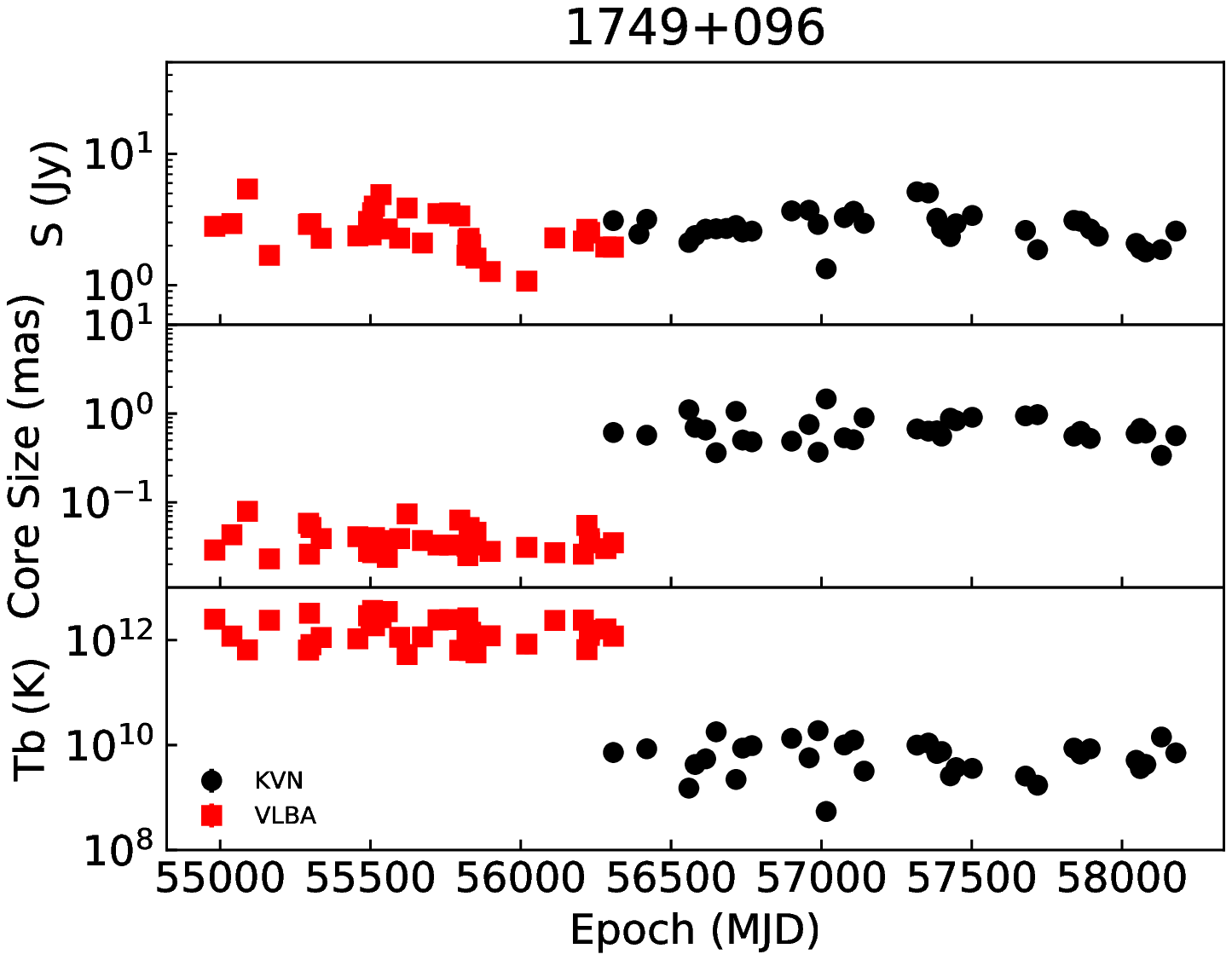}
\includegraphics[scale=0.4,trim={0.0cm 0cm 1.2cm 0cm},clip]{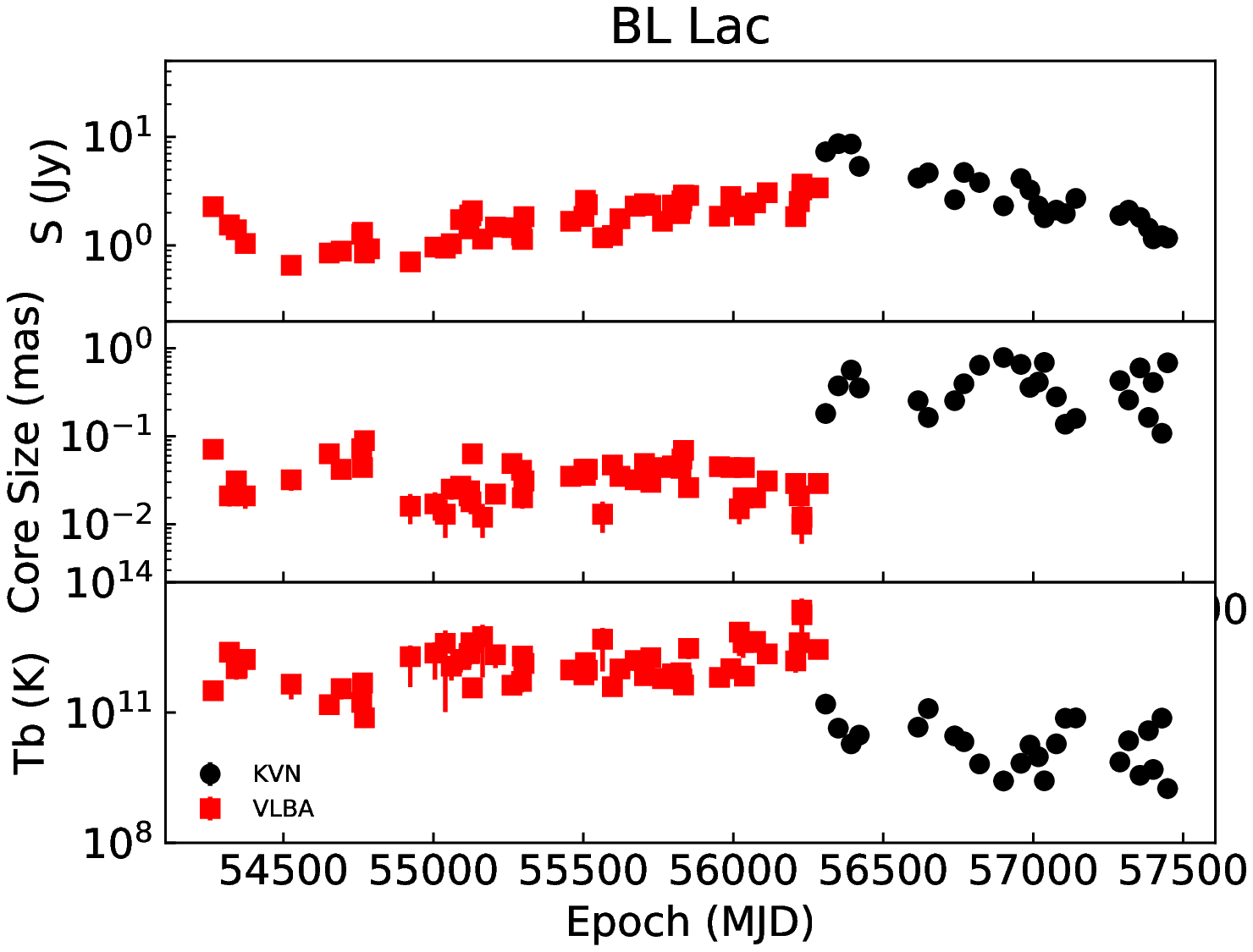}
\includegraphics[scale=0.4,trim={0.0cm 0cm 1.2cm 0cm},clip]{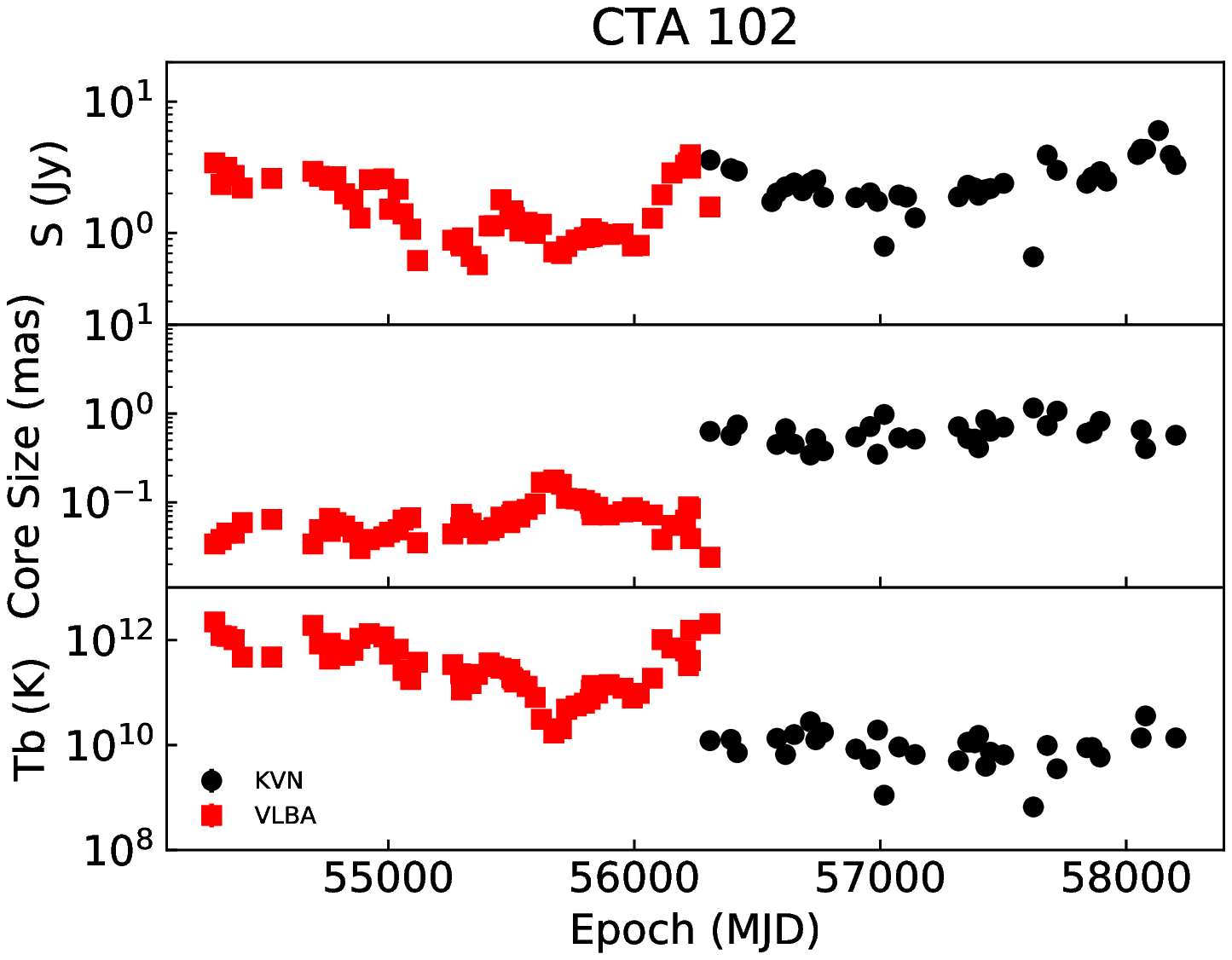}\\
\includegraphics[scale=0.4,trim={0.0cm 0cm 1.2cm 0cm},clip]{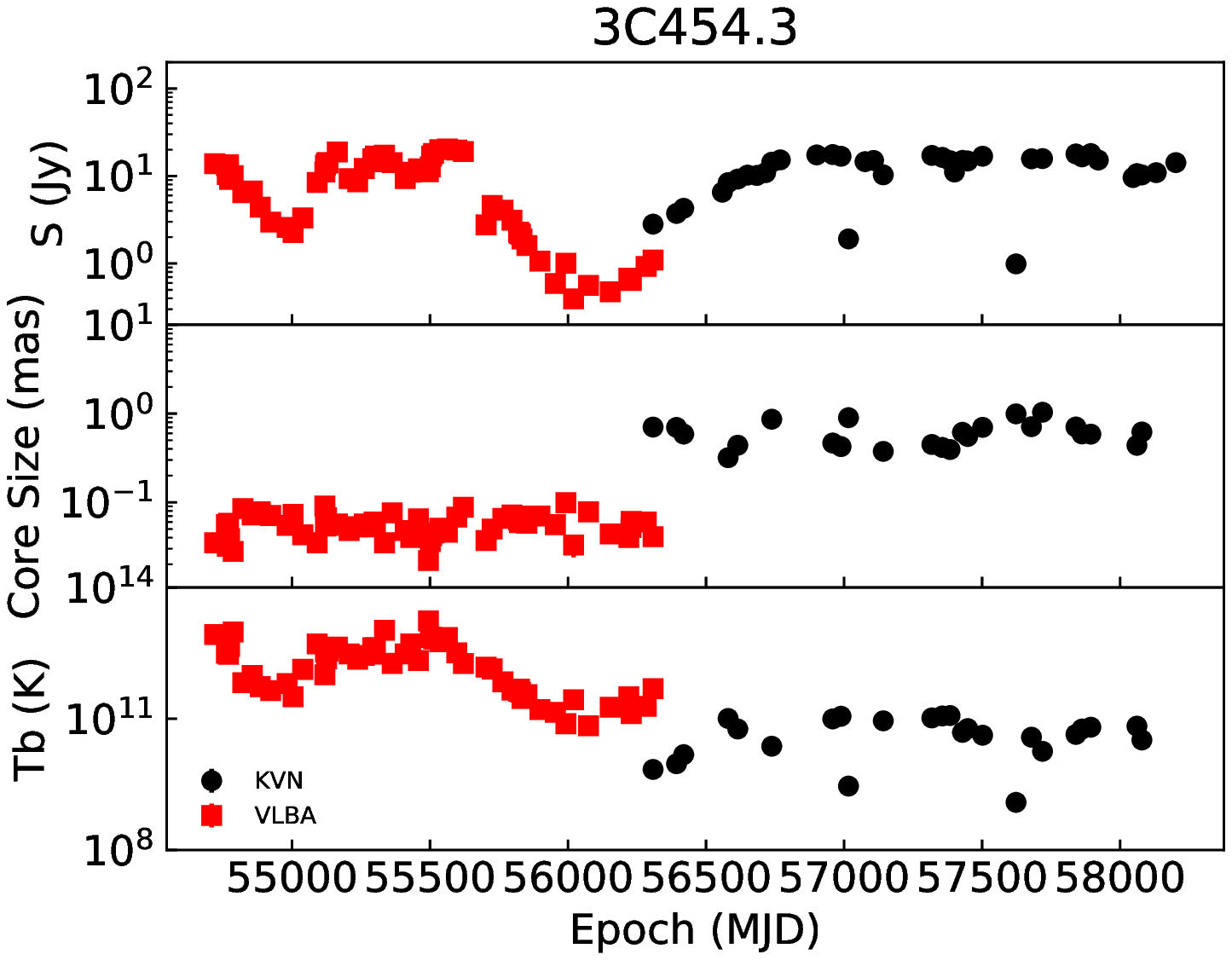}
\caption{--- Continued.} 
\end{figure*}

\begin{sidewaystable*}
\caption{Sources Median Quantities\label{table}}
\centering
\begin{tabular}{cccccccccccc}
\toprule
Source   & $z$ &  $\theta$($^{\circ}$) & $S^{VLBA}$(Jy)  &  $S^{KVN}$(Jy) & $f_S$ & $d^{VLBA}$(mas) & $d^{KVN}$(mas) &$f_d$ &$T_b^{VLBA}$($\times10^{10}$K) & $T_b^{KVN}$($\times10^{10}$K) & $f_{T_b}$\\
(1)&(2)&(3)&(4)&(5)&(6)&(7)&(8)&(9)&(10)&(11)&(12)\\
\midrule
0235+164 & 0.940 & 0.4 &$  1.08 \pm  0.95 $&$  1.31 \pm  0.51 $& 0.8  &$  0.05 \pm  0.02 $&$  0.60 \pm  0.26 $ & 0.09  &$ 32.00 \pm 38.66 $&$  0.39 \pm  0.45 $ & 81.5 \\
3C84 & 0.018 &39.1 &$  2.69 \pm  0.92 $&$  6.74 \pm 4.01 $& 0.4  &$  0.12 \pm  0.03 $&$  1.88 \pm  1.95 $ & 0.06  &$ 12.60 \pm 11.77 $&$  0.38 \pm  0.72 $ & 33.5 \\
3C111 & 0.048 & 15.5&$  0.83 \pm  1.17 $&$  1.27 \pm  1.22 $& 0.7  &$  0.06 \pm  0.04 $&$  0.70 \pm  0.23 $ & 0.09  &$ 18.40 \pm 19.34 $&$  0.17 \pm  0.54 $ & 106.8 \\
0420-014 & 0.915 & 1.9 &$  2.32 \pm  1.03 $&$  1.50 \pm  1.38 $& 1.6  &$  0.07 \pm  0.03 $&$  0.64 \pm  0.26 $ & 0.11  &$ 32.70 \pm 26.04 $&$  0.48 \pm  0.75 $ & 68.2 \\
0528+134 & 2.060 & 1.6 &$  1.56 \pm  1.28 $&$  0.94 \pm  0.20 $& 1.7  &$  0.04 \pm  0.02 $&$  0.66 \pm  0.20 $ & 0.06  &$ 54.65 \pm 87.93 $&$  0.45 \pm  0.31 $ & 121.2 \\
0716+714 & 0.127 & 5.2 &$  1.72 \pm  0.91 $&$  2.01 \pm  0.72 $&  0.9 &$  0.04 \pm  0.01 $&$  0.27 \pm  0.14 $ & 0.14  &$ 74.36 \pm 103.08 $&$  1.57 \pm  1.91 $ & 47.2 \\
0735+178 & 0.424 & ... &$  0.25 \pm  0.09 $&$  0.64 \pm  0.16 $& 0.4  &$  0.10 \pm  0.03 $&$  0.80 \pm  0.29 $ & 0.13  &$  1.77 \pm  1.08 $&$  0.08 \pm  0.13 $ & 21.5 \\
0827+243 & 0.939 & 3.9 &$  0.76 \pm  0.71 $&$  0.63 \pm  0.36 $& 1.2  &$  0.06 \pm  0.03 $&$  0.64 \pm  0.32 $ & 0.09  &$ 15.80 \pm 39.09 $&$  0.21 \pm  0.44 $ & 74.0 \\
0836+710 & 2.170 & 3.2 &$  1.40 \pm  0.73 $&$  1.36 \pm 0.45 $& 1.0  &$  0.07 \pm  0.02 $&$  1.11 \pm  1.29 $ & 0.06  &$ 19.15 \pm 15.77 $&$  0.23 \pm  0.34 $ & 84.8 \\
OJ287 & 0.306 & 3.3 &$  1.68 \pm  0.60 $&$  4.13 \pm  1.87 $& 0.4  &$  0.04 \pm  0.01 $&$  0.51 \pm  0.45 $ & 0.07  &$ 90.95 \pm 85.76 $&$  1.57 \pm  1.42 $ & 57.8 \\
1055+018 & 0.890 & 4.7 &$  2.40 \pm  0.57 $&$  2.22 \pm  0.71 $& 1.1  &$  0.05 \pm  0.01 $&$  0.36 \pm  0.07 $ & 0.13  &$ 84.75 \pm 65.23 $&$  2.39 \pm  0.26 $ & 35.4 \\
Mrk421 & 0.030 & ... &$  0.22 \pm  0.05 $&$  0.32 \pm  0.42 $& 0.7  &$  0.07 \pm  0.03 $&$  0.65 \pm  0.82 $ & 0.11  &$  3.15 \pm  2.65 $&$  0.05 \pm  0.06 $ & 61.1 \\
1156+295 & 0.729 & 2.0 & $ 0.96 \pm 0.63 $&$   1.15 \pm 0.64 $& 0.8  &$  0.04 \pm  0.02 $&$  0.54 \pm  0.24 $ & 0.08  &$ 55.11 \pm 77.20 $&$  0.61 \pm  1.68 $ & 90.0\\
1222+216 & 0.435 & 5.1 &$  0.77 \pm  0.45 $&$  1.48 \pm  0.48 $& 0.5  &$  0.04 \pm  0.02 $&$  1.20 \pm  0.19 $ & 0.03  &$ 46.90 \pm 37.55 $&$  0.10 \pm  0.06 $ & 453.1 \\
3C273B & 0.158 & 3.3 &$  3.14 \pm  2.27 $&$  8.58 \pm  4.19 $& 0.4  &$  0.08 \pm  0.04 $&$  1.01 \pm  0.42 $ & 0.08  &$ 33.15 \pm 77.33 $&$  0.74 \pm  1.01 $ & 44.7 \\
M87      & 0.004 & 14.0&$  0.70 \pm  0.00 $&$  1.40 \pm  0.15 $&   0.5 &$  0.11 \pm  0.01 $&$  0.54 \pm  0.11 $ & 0.20  &$  3.83 \pm  0.86 $&$  0.31 \pm  0.15 $ & 12.5 \\
1308+326 & 0.996 & 3.2 &$  1.04 \pm  0.66 $&$  1.01 \pm  0.37 $& 1.0  &$  0.07 \pm  0.02 $&$  0.70 \pm  0.27 $ & 0.10  &$ 15.80 \pm 10.28 $&$  0.21 \pm  0.31 $ & 74.4 \\
1510-089 & 0.361 & 3.4 &$  1.63 \pm  0.78 $&$  2.57 \pm  1.41 $& 0.6  &$  0.06 \pm  0.04 $&$  0.89 \pm  0.54 $ & 0.07  &$ 30.50 \pm 55.44 $&$  0.35 \pm  0.66 $ & 87.2 \\
1633+382 & 1.813 & 2.5 &$  2.10 \pm  0.75 $&$  3.41 \pm  1.65 $&   0.6 &$  0.05 \pm  0.03 $&$  0.44 \pm  0.27 $ & 0.11  &$ 158.65 \pm 276.18 $&$  3.82 \pm  7.10 $ & 41.5 \\
3C345 & 0.595 & 5.1 &$  1.40 \pm  0.39 $&$  3.46 \pm  0.94 $& 0.4  &$  0.05 \pm  0.01 $&$  0.50 \pm  0.23 $ & 0.10  &$ 45.15 \pm 43.99 $&$  1.34 \pm  1.05 $ & 33.7 \\
NRAO530 & 0.902 & 3.0 &$  1.90 \pm  0.82 $&$  3.17 \pm  0.69 $& 0.6  &$  0.05 \pm  0.02 $&$  1.02 \pm  0.30 $ & 0.05  &$ 50.10 \pm 59.91 $&$  0.41 \pm  0.21 $ & 121.0 \\
1749+096 & 0.322 & 3.8 &$  2.50 \pm  0.93 $&$  2.69 \pm  0.79 $& 0.9  &$  0.04 \pm  0.01 $&$  0.63 \pm  0.24 $ & 0.06  &$ 124.00 \pm 91.75 $&$  0.69 \pm  0.46 $ & 179.4 \\
BL Lac    & 0.069 & 7.3 &$  1.84 \pm  0.74 $&$  2.48 \pm  2.21 $&   0.7 &$  0.03 \pm  0.02 $&$  0.37 \pm  0.20 $ & 0.09  &$ 126.30 \pm 390.97 $&$  2.02 \pm  4.01 $ & 62.5 \\
CTA102 & 1.037 & 3.7 &$  1.31 \pm  0.90 $&$  2.35 \pm  1.04 $& 0.6  &$  0.06 \pm  0.03 $&$  0.60 \pm  0.20 $ & 0.10  &$ 30.25 \pm 51.59 $&$  0.92 \pm  0.73 $ & 32.9 \\
3C454.3 & 0.859 & 1.3 &$  8.87 \pm  6.51 $&$ 14.40 \pm  4.88 $& 0.6  &$  0.06 \pm  0.02 $&$  0.59 \pm  0.20 $ & 0.09  &$ 184.50 \pm 334.00 $&$  5.36 \pm  3.87 $ & 34.4 \\
\bottomrule
\end{tabular}

\vspace{1.5cm}

\caption{Selected Sources Quasi-Simultaneous Median Quantities\label{newtable}}
\centering
\begin{tabular}{cccccccccccc}
\toprule
Source   & $z$ &  $\theta$($^{\circ}$) & $S^{VLBA}$(Jy)  &  $S^{KVN}$(Jy) & $f_S$ & $d^{VLBA}$(mas) & $d^{KVN}$(mas) &$f_d$ &$T_b^{VLBA}$($\times10^{10}$K) & $T_b^{KVN}$($\times10^{10}$K) & $f_{T_b}$\\
(1)&(2)&(3)&(4)&(5)&(6)&(7)&(8)&(9)&(10)&(11)&(12)\\
\midrule
0716+714 & 0.127 & 5.2 &$  1.72 \pm  0.91 $&$  2.01 \pm  0.72 $&  0.9 &$  0.04 \pm  0.02 $&$  0.30 \pm  0.14 $ & 0.11  &$ 90 \pm 103.08 $&$  1.30 \pm  1.91 $ & 68.2 \\
1156+295 & 0.729 & 2.0 & $ 0.96 \pm 0.63 $&$   1.15 \pm 0.64 $& 0.8  &$  0.04 \pm  0.02 $&$  0.49 \pm  0.26 $ & 0.08  &$ 330 \pm 77.20 $&$  0.48 \pm  1.68 $ & 136.0\\
1633+382 & 1.813 & 2.5 &$  2.10 \pm  0.75 $&$  3.41 \pm  1.65 $&   0.7 &$  0.05 \pm  0.01 $&$  0.44 \pm  0.27 $ & 0.09  &$ 660 \pm 276.18 $&$  3.82 \pm  7.10 $ & 86.5 \\
\bottomrule
\end{tabular}

\end{sidewaystable*}

\begin{table}
\caption{Sample Fractional Quantities\label{medianfractional}}
\centering
\begin{tabular}{cccc}
\toprule
& $f_S$ &  $f_d$   &  $f_{T_b}$\\
\midrule
Minimum &        0.4 & 0.03&  12.5  \\
25\% Quartile &      0.5  &  0.07&  36.9\\
50\% Quartile  &     0.6   & 0.09&  61.8\\
75\% Quartile &     0.9   & 0.11&  86.6\\
Maximum   &     1.6   & 0.20& 453.1\\
\bottomrule
\end{tabular}
\end{table}

\subsection{Notes on Individual Sources}

Below, we indicate some remarks for a number of sources for which the assumptions provided above (small variability, etc) may not apply. In effect, these sources may provide some bias for the discussion, and care should be taken when considering their observed quantities and derived fractional values.

-- \emph{3C84:} Significant amount of upper limits were found for the core size in this source, suggesting that the actual core is much smaller than what the KVN can resolve at 43~GHz. As a consequence, no brightness temperatures were calculated for such upper limits. 

-- \emph{0420-014:} The compactness factor is $f_S>1$ for this source. However, the observed flux with VLBA seems to connect well with the one with KVN. We consider that this larger value may be rather caused by variability effects.

-- \emph{0528+134:} The observed VLBA flux of this source appears to be larger than its KVN flux. Given that the resolution of KVN is smaller and, a priori, the core flux should include a larger region, possibly leading to larger flux densities, if any, we consider the large compactness factor $f_S>1$ for this source to be also due to variability effects.

-- \emph{OJ287:} The flux density of this source seems to be steadily increasing. As a consequence, even when the VLBA flux smoothly connects with the KVN flux, suggesting a ratio close to unity, the actual VLBA and KVN median values appear to be different, leading to a fiducial compactness=0.4

-- \emph{1222+216:} The flux density seems to be significantly variable in this source, with two clear maxima in the data. There is a local minimum located in the VLBA data, which may bias the compactness of this source to a value lower than the one which would be expected. 

-- \emph{3C454.3:} Significant flux variability is found in this source for the VLBA data, with several minima found. The data for KVN suggests a more stable flux density after a small increase.

\section{Discussion}

As expected, due to the comparatively poorer resolution of the KVN, both the observed flux densities and core sizes appear to be larger than these from the VLBA. In the case of the flux densities, the factor $f_S$ provides information about the compactness of the source. Given an extended structure, integration of the flux over a larger region will result in a larger value, but not all of such flux will actually arise from the more unresolved region observed within the larger array. In our case, $f_S\sim0.6$ suggests that, on average, VLBA can observe roughly only 60\% of the KVN flux or, inversely, 40\% of the flux considered to be arising from the VLBI KVN core may be emitted in other regions \citep[although a proper study using convolved images should confirm this; see][]{Kim19}. Note however that this value changes dramatically with different sources,  as shown by the various quartiles. In some sources (e.g., 3C84, 0735+178, OJ287, 3C273B, 3C345), more than half of the core KVN flux can be attributed to blending effects, whereas in more compact ones (e.g., 0716+714, 0836+710, 1308+326) most of the KVN flux seems to arise from the core regions. This is in agreement with the discussion in \cite{Rioja14}, where they suggest that the magnitude of the blending strongly depends on the source.

Maximum baselines for the case of VLBA are of the order of 8611~km (between Mauna-Kea and Saint-Croix antennas), leading to resolutions of the order of 0.17~mas at 43~GHz. On the other hand, maximum KVN baselines are of the order of 476~km (between Tamna and Yonsei antennas), leading to resolution of about 3.0~mas. This is an improvement of the resolution by a factor of 18 when using the VLBA array for 43~GHz. We note however that the maximum baselines (and hence, the smaller beam sizes) will only occur in the ideal cases, and the actual baselines will be given by several other factors such as source elevation.  In order to investigate this in detail, we have performed a check as follows: we obtained typical beam sizes from VLBA data \citep{Jorstad17} and iMOGABA data \citep{Lee16} and obtained the fraction $f_{beam}$=beam(VLBA)/beam(KVN) on a source-by-source basis. Once we consider the more realistic beam sizes, rather than the maximum baselines, the factor in resolution between VLBA and KVN become  $f_d\sim1/10$, which is much closer to the 50\% quartile for $f_d$. We thus consider that, although they may have a slightly effect, the baseline length is not significantly affecting our results.

There is however no significant correlation found between $f_{beam}$ and $f_S$ (Pearson $r=0.07$), and only moderate for $f_{beam}$ and $f_d$ ($r=0.47$).  Additionally, the normalized standard deviation $\sigma(f_{beam}/f_{beam}^{max})=0.07$ is much smaller than the normalized standard deviation for $f_S$ and $f_d$ (0.23, 0.21, respectively). We have checked that the beam size for the BU-VLBA observations can change by a factor of $\lesssim11$\%. Similarly, the beam size of the beam size for iMOGABA observations change by a factor $\lesssim10$\%. This is smaller than the dispersion that we find in the factors $f_S$ and $f_d$, which can vary by about an order of magnitude. This suggests that the dispersion found in the values for $f_S$ and $f_d$ may have a different origin. 

The derived value for the brightness temperature also get severely affected as a consequence of the two factors $f_S$ and $f_d$. In principle, as we use a larger array, we should be able to probe smaller regions. At the same time, if the source were uniform, we would also observe smaller flux densities in the proportion $S\propto d^2$, thus leading to a similar $T_b$. However, this is clearly not the case in AGNs in general, with a brighter core and blending effects. In our case here, on average VLBA fluxes are 60\% smaller but sizes values become only 9\% of their KVN value. This large disproportion derives in brightness temperatures $10^{1-2.5}$ times larger with the VLBA array.

\subsection{Origin of the Dispersion in $f_S$ and $f_d$ }

The dispersion that we find in the fractional values seems to be quite significant and cannot be attributed only to the measurement uncertainties. Knowing what is the origin of such large dispersion is crucial to asses a proper factor in accounting for core properties with the KVN. It is possible that the source itself plays an important role due to blending effects.

\begin{figure}[ht]
\center
\includegraphics[scale=0.58,trim={0cm 0cm 0cm 1cm},clip]{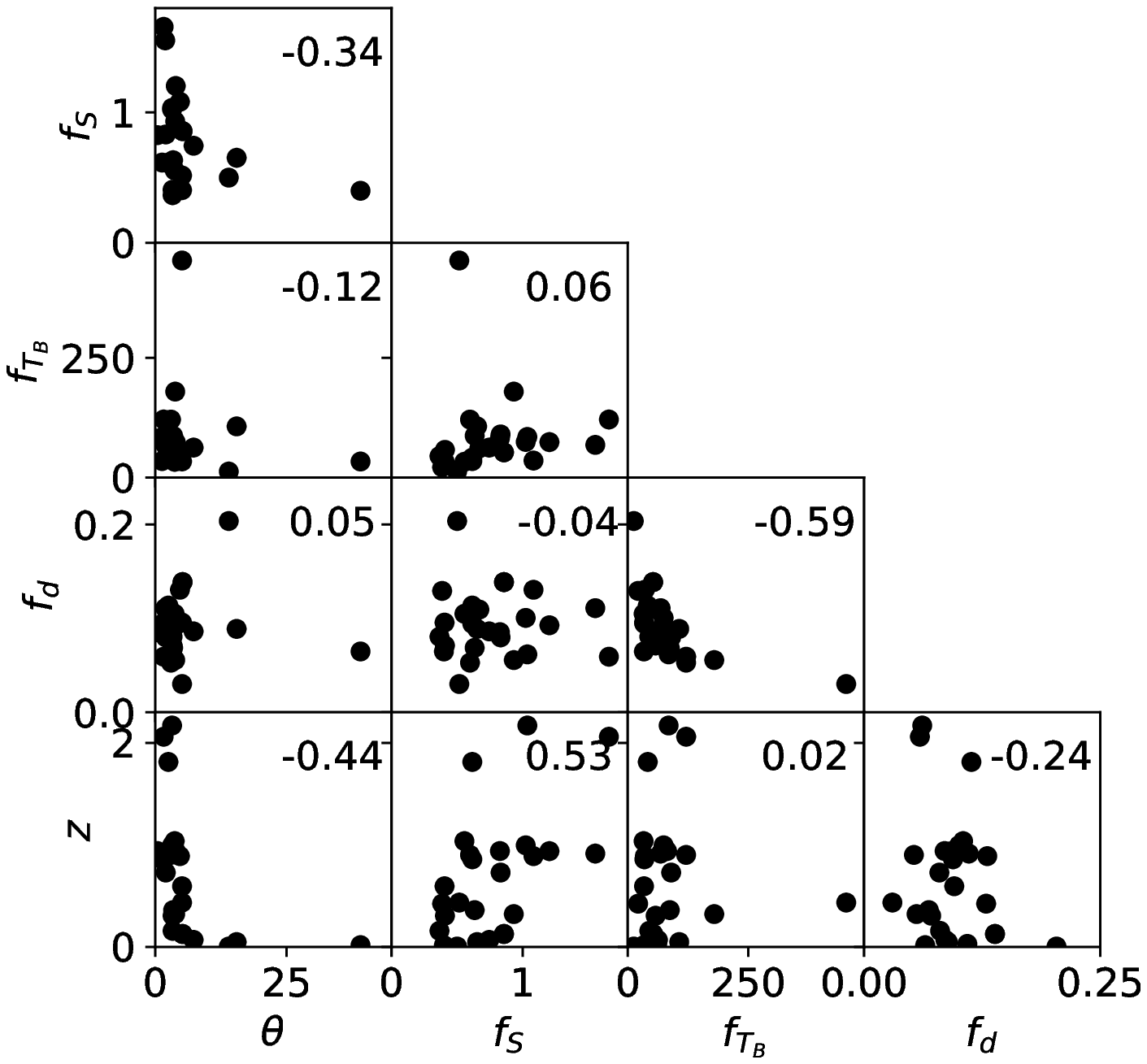}
\caption{Scatter matrix for the proposed correlations. The correlation coefficients are shown in the top right corner of each panel.} 
\label{fig_combinations}
\end{figure}

On one hand, the derived parameters for each source may be intrinsically different due to i) compactness (an increase of the resolution would not significantly affect the observables of a compact source), ii) small viewing angles (leading to components appearing closer in projection), or iii) different redshift (features would appear smaller). Based on this, we consider any possible dependence on the size and brightness temperature factors $f_d$ and $f_{T_b}$ in terms of the compactness $f_S$, redshift $z$ and viewing angle $\theta$ of the sources. In Figure \ref{fig_combinations} we show the proposed correlations, including the Pearson correlation coefficient $r$. Inspection of the figure clearly shows that there seems to be no relevant dependence of the blending with compactness, redshift or viewing angle\footnote{Note that some fiducial correlation appears due to observational bias ($\theta$ vs $z$) or parameters dependence ($f_{T_B}$ vs $f_d$).  }.

Alternatively, since blazars are highly variable in structure and in flux, the blending effect is highly time-dependent. Such variability may produce changes, or even spurious values, in the measured fractional parameters $f_S$ and $f_d$ if comparison is not made properly. Since, the two data sets, KVN and VLBA, used here are not fully coincident in time, variability effects may play a major role. Indeed, although we considered observations over a long period of time to circumvent the need for quasi-simultaneous observations, we found that for all sources, the flux density standard deviation was larger than 15\% the median flux, suggesting a significant variability. Indeed, we note that i) the most extreme value $f_d=0.03$ corresponds to 1222+216, which has been noted to have an observational bias and ii) the sources containing quasi-simultaneous VLBA and KVN data (0716+714, 1156+295, and 1633+382) seem to have similar $f_d$ values. Thus, data analysis of non-overlapping time may have introduced uncertainties and probably biases.

Reflecting this, we consider the sources above mentioned, 0716+714, 1156+295 and 1633+382, and focus on the epochs which have KVN and VLBA data overlapped in time. In Table \ref{newtable} we summarize the new quasi-simultaneous median quantities for these sources. It seems that the values of $f_d$ for these sources, which were already close, become more similar when not only all the data range but only quasi-simultaneous data is considered. As a different check, we also considered the effect of removing the quasi-simultaneous data for these sources and performed various tests flagging VLBA data near in time to that of the KVN (for example, flagging VLBA data after MJD$>56000$, as a simple case). We found that the values for $f_d$ significantly changed, with cases where $f_d=0.19$ for 1633+382, or $f_d=0.14$ for 0716+714. This suggests that variability effects are indeed the source for the dispersion in the  $f_d$ values. We thus suggest a blending factor for the KVN of $f_d\sim0.09$.

It seems, on the other hand, that $f_S$ is intrinsically source-dependent, and there doesn't seem to be a simple recipe to consider a priori this parameter. Not only the compactness of the source but a full analysis of its structure should be taken into account. Furthermore,  ejection of new components, possibly associated with $\gamma-$ray flares, may alter the innermost structure of the source, and a more methodic study, beyond the scope of this work, should follow \cite[see e.g.][]{Rioja14}. Additionally, given its dependence on the flux, it will not be straightforward to find a common factor for $f_{T_B}$ either.

\subsection{Extrapolation to Other Frequencies}

Regarding the core size, it seems that the factor $f_d=0.09$ considered here seems to be in agreement with the value expected considering the array resolution, once we consider the actual baselines and UV coverage during the observations. It is thus reasonable to consider that, at different frequencies, this factor will be scaled accordingly. On the other hand, given the characteristics for $f_S$ and hence $f_{T_B}$, considerations for the brightness temperature may not be straightforward.

In \cite{Lee14}, it is discussed that, in general for AGNs, $T_b \propto \nu^{\xi}$, with $\xi=+2.6$ below a critical frequency $\nu_c$, which corresponds to the peak frequency of the spectrum. Beyond this frequency, $\xi\sim -1$ for a decelerating jet model and $\xi\sim +1$ for the rapidly accelerating jet model. In that work, it was found that the brightness temperature seemed to decrease with frequency as $T_b \propto \nu^{-1.2}$ for $\nu > 9$~GHz, favoring the decelerating jet model. However, in \cite{Lee16}, the median brightness temperatures increase from $T_b=10^9~K$ at 22~GHz to $T_b=7.4\times 10^9~K$ at 129~GHz; i.e, increasing by almost an order of magnitude. Furthermore, the observed frequency dependence in \cite{Lee14} was significantly different than the predictions. These apparent inconsistencies could be potentially due to the blending effects discussed here. Only once these are understood, the physical model can be truly tested.

From our result above, we can consider that, statistically speaking, the actual brightness temperature at 43~GHz may be a factor $\sim50$ larger than the one observed with KVN. If we assume an accelerating jet, it may be possible for the blending effects to be similar or larger at higher frequencies. However, under the assumption of a constant speed or decelerating jet, blending effects on the KVN should decrease at higher frequencies.

\section{Conclusions}

We investigate the effects of core blending effects on AGNs under the KVN view by comparing the properties of a sample of 25 sources when observed with the KVN and VLBA arrays. For this purpose, we collected data at 43~GHz from the KVN iMOGABA program and the  43~GHz BU-VLBA-Blazar program and supplemented it with some additional observations. Although the two data sets are not fully coincident in time, we consider various cases where quasi-simultaneous observations exist and study their effects on the discussed quantitites.

Our results suggest that, on average, the core flux densities are larger by a factor $f_S=0.6$; the core sizes are larger by a factor $f_d=0.09$, and the brightness temperatures are lower by a factor $f_{T_b}=59$, when observed with the KVN. These factors are compatible with the a priori expectations purely based on the arrays different resolutions. Note however that, although a common blending factor $f_d$ would suffice to characterize the KVN with respect to other VLBI arrays, there is a significant scatter in the fractional values for the flux density $f_S$ and the brightness temperature $f_{T_b}=59$. Such scatter may be attributed to the particular properties of each source, as suggested by previous results by \cite{Rioja14}.

We thus suggest that a factor $f_d=0.09$ could be used to scrutinize KVN core size blending effects when comparing the VLBA and KVN at 43GHz. Otherwise, a source--dependent factor can also be estimated. We discuss considerations regarding the AGN jet model and possible implications of the relative magnitude of the blending effect at different frequency bands. Further work, including simulations and matching spatial resolutions of the observations will be the topic of future research.

\acknowledgments
We are grateful to all staff members in KVN who helped to operate the array and to correlate the data. The KVN is a facility operated by the Korea Astronomy and Space Science Institute. The KVN operations are supported by KREONET (Korea Research Environment Open NETwork) which is managed and operated by KISTI (Korea Institute of Science and Technology Information). We acknowledge financial support from the Korean National Research Foundation (NRF) via Global Ph.D. Fellowship Grant 2014H1A2A1018695 and Basic Research Grant NRF-2015R1D1A1A01056807. S. S. Lee was supported by the National Research Foundation of Korea (NRF) grant funded by the Korea government (MSIP) (NRF-2016R1C1B2006697). J. W. Lee is grateful for the support of the National Research Council of Science and Technology, Korea(Project Number EU-16-001). This study makes use of 43 GHz VLBA data from the VLBA-BU Blazar Monitoring Program (VLBA-BU-BLAZAR; http://www.bu.edu/blazars/VLBAproject.html), funded by NASA through the Fermi Guest Investigator Program. The VLBA is an instrument of the National Radio Astronomy Observatory. The National Radio Astronomy Observatory is a facility of the National Science Foundation operated by Associated Universities, Inc.

\end{document}